\documentclass[aps,prb,twocolumn,showpacs,superscriptaddress,amssymb]{revtex4}
\usepackage{graphicx}
\usepackage{amsmath}
\usepackage[usenames]{color}
\usepackage{hyperref}
\usepackage{natbib}
\usepackage{bm}
\usepackage{threeparttable}
\def\Tr{\text{Tr}}

\begin{document}

\title{Comparison of dynamical decoupling protocols for a nitrogen-vacancy center in diamond}

\author{Zhi-Hui Wang}
\affiliation{Ames Laboratory, Iowa State University, Ames, IA, 50011}
\author{G. de Lange}
\affiliation{Kavli Institute of Nanoscience Delft, Delft University of Technology,
P.O. Box 5046, 2600 GA Delft, The Netherlands}
\author{D. Rist\`e}
\affiliation{Kavli Institute of Nanoscience Delft, Delft University of Technology,
P.O. Box 5046, 2600 GA Delft, The Netherlands}
\author{R. Hanson}
\affiliation{Kavli Institute of Nanoscience Delft, Delft University of Technology,
P.O. Box 5046, 2600 GA Delft, The Netherlands}
\author{V. V. Dobrovitski}
\affiliation{Ames Laboratory, Iowa State University, Ames, IA, 50011}

\date{\today}
\begin{abstract}
We perform a detailed theoretical-experimental study of the
dynamical decoupling (DD) of the nitrogen-vacancy (NV) center in diamond.
We investigate the DD sequences applied to suppress the dephasing of the electron spin of the NV center induced by the coupling to a spin bath composed of the substitutional nitrogen atoms. The decoupling efficiency of various DD schemes is studied, including both periodic and aperiodic pulse sequences.
For ideal control pulses, we find that the DD protocols with the Carr-Purcell-Meiboom-Gill (CPMG) timing of the pulses provides best performance. We show that, as the number of control pulses increases, the decoupling fidelity scaling differs qualitatively from the predictions of the Magnus expansion, and explain the origin of this difference. In particular, more advanced symmetrized or concatenated protocols do not improve the DD performance. Next, we investigate the impact of the systematic instrumental pulse errors in different periodic and aperiodic pulse sequences. The DD protocols with the single-axis control do not preserve all spin components in the presence of the pulse errors, and the two-axis control is needed. We demonstrate that the two-axis control sequence with the CPMG timing is very robust with respect to the pulse errors. The impact of the pulse errors can be diminished further by symmetrizing this protocol. For all protocols studied here, we present a detailed account of the pulse error parameters which make strongest impact on the DD performance. In conclusion, we give specific recommendations about choosing the decoupling protocol for the system under investigation.
\end{abstract}

\maketitle

\section{Introduction}
A singly negatively charged nitrogen-vacancy (NV) center in diamond
has recently emerged as a promising candidate
for solid-state quantum computation and
quantum information processing,\cite{JelezkoGate04,Dutt07,Childress06PRL,Cappellaro09,Jiang09,Neumann10,Buckley10,Fuchs11,Togan10,Robledo11} and high-precision electric and magnetic sensor for nanoscale applications.\cite{Taylor08,Maze08,Balasubramanian08,deLangeMagnetometry10,McGuinness11,Cole09,Hall09,Dolde11,Meriles10,Budker11}
The quantum state of the electron spin of a single NV center can be
conveniently initialized and read-out optically,\cite{Gruber97,Jelezko02,Harrison04}
and coherently manipulated electrically, optically, and magnetically, even at room temperatures.\cite{Jelezko04,Santori06,Hanson08,Childress06,Gaebel06,Fuchs09,Fuchs10,Tamarat06} By prolonging the coherence time of the NV center spin,
these favorable properties can be better exploited for potential applications.

Dynamical decoupling (DD) is an efficient tool for decoherence suppression. In nuclear magnetic resonance, the techniques based on Hahn spin echo, which employ the control pulses to manipulate the spins and achieve high-resolution spectra, have been widely used for several decades.\cite{Haeberlen}
Lately, DD has being actively explored in the general context of QIP. By applying a sequence of control pulses ($\pi$-rotations) to the qubit, in the limit of very short inter-pulse delay,
DD can effectively cancel the coupling between the system and its decohering environment, thus preserving the qubit coherence.\cite{Viola99,Viola98} Various DD schemes have been developed,\cite{Viola99,Viola98,Santos05,Khodjasteh07,Uhrig07,UhrigCUDD09,Biercuk09,Uys09,West10} and implemented on different spin systems.\cite{Fraval05,Biercuk09,Uys09,Bluhm10,Morton08,Tyryshkin10,Morton06,Du09,MarcusDDQD}

DD has been recently implemented
on single NV centers.\cite{Naydenov10,deLange10,Ryan10} Excellent performance of several basic DD sequences has been demonstrated, thus opening the way to using DD in various experimental applications of NV centers. Understanding the performance of different DD schemes on NV centers in diamond is timely, and will be useful for future developments in this area of research. In this paper, we present a detailed study of various DD sequences designed to preserve the quantum state of a single NV center spin decohered by a spin bath of substitutional nitrogen defects (P1 centers), which is the dominant decoherence source for the type Ib diamonds. We identify the most useful DD protocols, and study their efficiency in realistic experimental situations.

A large family of DD protocols is constructed based on the pulse sequences with periodic structure. Representatives are Carr-Purcell-Meiboom-Gill (CPMG) sequence and its generalizations,\cite{Slichter,Gullion90}
and periodic dynamical decoupling (PDD).\cite{Viola98,Viola99,Santos05,Zhang07}
Among them, the sequences where all pulses rotate the spin about the same axis (single-axis control) refocus the spin dephasing,
while the sequences where the spin is rotated alternatively
about two mutually orthogonal axes (two-axis control)
are able to suppress general decoherence.\cite{Viola99}
However, the inter-pulse delay can never be made arbitrarily small to completely eliminate decoherence,
because of unavoidable experimental constraints. These constraints limit the efficiency of the basic DD schemes.
To suppress the effect of system-environment coupling more efficiently, more advanced symmetrized versions of periodic dynamical decoupling,\cite{Haeberlen,ViolaEDD}
and the concatenated dynamical decoupling (CDD) have been proposed.\cite{Khodjasteh05,Khodjasteh07}

Another important family of DD protocols is based
on aperiodic DD schemes, such as Uhrig's DD (UDD),
where the pulse spacings are optimized to suppress
dephasing with a given number of pulses.\cite{Uhrig07}
A DD scheme extending UDD to the two-axis control, the quadratic DD (QDD)
has been proposed for suppression of general decoherence.\cite{West10}
The performance of a given decoupling sequence strongly depends on the nature of the decohering bath.
It has been demonstrated that
for a bath with rapidly decaying spectral density,
UDD performs better
than periodic protocols (like CPMG) with the same number of pulses,
while for a bath with a soft frequency cutoff in the spectral density,
CPMG-like pulse spacing is preferred.\cite{Cywinski08,Cywinski09,WYang08,Biercuk09,Pasini10,deLange10,Ryan10,Suter10} In a more general situation, the pulse positions can be numerically optimized for a given type of the bath.\cite{Biercuk09,Uys09,Biercuk09PRA,Hayes11}

Here we consider the decohering bath made of the electron spins of P1 centers, which are coupled to the NV center and to each other via long-range dipolar coupling. The effect of such a bath is accurately approximated by the mean-field model of a random magnetic field acting on the NV center and leading to the dephasing of the NV spin. \cite{Hanson08,deLange10} We perform theoretical and experimental study of DD in such a system, studying both periodic and aperiodic DD sequences based on single- and two-axis control, as well as their symmetrized and concatenated versions. Using both analytical and numerical tools, we show that the periodic sequences with equally spaced pulses (PDD) and the sequences with CPMG timing exhibit different short and long time behavior. The aperiodical sequences UDD and QDD show inferior performance in comparison with the CPMG-timed sequences, in accordance with the previous studies mentioned above.
We further demonstrate that, in contrast with the expectations, the concatenated DD sequences do not improve the performance of decoupling; we clarify the origin of this behavior.

For ideal DD pulses, the decoupling efficiency grows in the limit of very short inter-pulse delay (in non-pathological cases). On the other hand, as the number of DD pulses is increased to achieve shorter delay, the negative effect of the unavoidable imperfections in the control pulses becomes noticeable. At long times, the pulse error accumulation can completely destroy the DD performance,\cite{Slichter,GersteinBook} and understanding the impact of the pulse errors is important for successful decoupling. Below, we focus on the instrumental pulse errors caused by imperfect adjustment of the rotation axis/angle (the errors caused by decoherence during the control pulse are much less important in our experiments). Since 1970s, a vast number of recipes has been suggested to reduce the effect of the instrumental pulse errors, from pulse tuning and shaping  minimizing the individual pulse errors, \cite{BurumEtal81,ShakaEtal88,FortunatoCoryPulses,CoryPulses1,Khaneja05,Dobrovitski10,MortonSPAM,Sengupta05} to combining the imperfect pulses in order to compensate the errors.\cite{Levitt,ShakaKeelerReview,ViolaEDD,Suter11} In our work, minimization of the errors is not addressed: we assume them fixed. We analyze how different sequences accumulate or compensate these fixed instrumental pulse errors: this is known to drastically depend on the specific sequence.\cite{Khodjasteh05,Khodjasteh07,Biercuk09,Biercuk09PRA,Suter10,Tyryshkin10,Wang10,WangJOP11,KhodjPSE} We show that the CPMG and the UDD sequences accumulate pulse errors in the rotation angle and the $z$-component of the rotation axis, with the decoupling fidelity being sensitive to the initial spin state.
The sequences based on the two-axis control are in general more robust to the pulse errors; we provide a detailed account of the most important components of the pulse errors. As a result, our results provide a detailed understanding of the dynamical decoupling of a single NV center spin in the bath of P1 centers in diamond.

The rest of this paper is organized as follows:
In Sec~\ref{sec:system} we describe the system under study, a NV center in diamond. Sec~\ref{sec:exp} presents description of the experiments. In Sec~\ref{sec:scaling} we analyze the decoupling performance
of the DD protocols assuming ideal pulses.
In Sec~\ref{sec:err} we focus on the accumulation of
pulse errors in various DD sequences for different states of the NV center's spin.
In Sec.~VI we discuss the internal structure of the bath under study. Conclusions are given in Sec.~VII.

\section{Description of the system: a NV center in diamond}
\label{sec:system}
The singly negatively charged NV center in diamond is composed of a vacancy defect and an adjacent substitutional nitrogen atom.
The electronic ground state of the center has a total spin $S=1$ with a zero-field splitting $D=2.87~\text{GHz}$ between the $m_S=0$ and $m_S=\pm 1$ levels.\cite{Wrachtrup06}
The quantization axis of this splitting
is along the symmetry axis of the NV center, which we take as the $z$ axis.
In our experiments, a static magnetic field $B_0=114$~G is applied along the $z$ axis, and the sublevels $m_S=+1$ and $m_S=-1$ are separated.
The transition energies between states $m_\text S=0$ and  $m_\text S=+1$
and that between $m_\text S=0$ and  $m_\text S=-1$ differ by hundreds of MHz.
We apply pulses to the NV center only in resonance with the transition between $m_\text S=0$ and  $m_\text S=-1$,
which can thus be regarded as an effective two-level system $S_0=1/2$ (the central spin).
The NV center spin is coupled to the nuclear spin of the host nitrogen atom, $I_0$, via hyperfine interaction $A_0S^z_0I^z_0$, where $A_0=-2\pi\cdot 2.16$~MHz for $^{14}$N, and $A_0=2\pi\cdot 3.03$~MHz for $^{15}$N nuclei, respectively. \cite{ChildressHF}
The intrinsic relaxation time of the nuclear spin
is of orders of milliseconds, so that $I_0$ for a single experimental run
is a constant of motion which merely shifts the Larmor frequency of the NV center spin, but changes from one run to another.

The NV center is surrounded by a large number of P1 centers (abundance $\sim0.02$~\% in our sample), which form the decohering bath.
Each P1 center is composed of an unpaired electron with spin $S_k=1/2$
and a $^{14}\text N$ nuclear spin $I_k=1$, coupled via hyperfine interation of the order of $100$~MHz.
The NV center spin $S_0$ is dipolarly coupled to all $S_k$ with the coupling strength $\sim 1$~MHz, and the dipolar coupling between the electron spins of different P1 centers is also of order of 1~MHz.
The dipolar interactions between different spins $S_k$, therefore,
are responsible for the internal dynamics of the spin bath, leading to the flip-flops between different bath spins.
At the same time, the flip-flops between the central spin $S_0$ and the bath spins $S_k$ are strongly suppressed,
due to the large mismatch in the transition energies (2.55~GHz for the
NV center spin vs.\ a few hundreds of MHz for a P1 center).
Therefore, decoherence of the NV center spin is dominated by pure dephasing, i.e.\ decay of the transversal (the $x$-$y$ plane) component of the NV spin.
The longitudinal $z$-component, which decays at much longer times (tens of milliseconds and up, mostly due to spin-phonon relaxation), is assumed to be conserved. In this situation, the bath of P1 centers can be represented \cite{Hanson08,deLange10} as a bath of spins 1/2 (which we denote as $\tilde S_k$), with the system-bath coupling Hamiltonian
\begin{equation}
H_{SB}=S^z_0\sum_k a_k \tilde S^z_k
\end{equation}
Each coupling constants $a_k$ depends on the specific state of the nuclear
spin of the $k$-th P1 center, on the direction of the symmetry axis of this
P1 center, and on the location of this P1 center with respect to the NV center.
For details, see Sec.~\ref{sec:bathlines} below, and the supporting material of Ref.~\onlinecite{Hanson08}.

The decoherence problem for a central spin coupled to a many-spin bath
is typically beyond the reach of analytical treatment due to the complex intra-bath dynamics. Many methods have been developed in the field of magnetic resonance,\cite{KlauderAnderson,Kubo1,HuHartmann,Salikhov81} and recently in the field of quantum information processing.\cite{deSousa03,Witzel05,RBLiu07,Saikin07,WYang08,Dobrovitski08,Hanson08,Cywinski09,Cywinski10,Cywinski11,Maze08b,Cappellaro09,Dobrovitski09,CoishBaugh,CoishLoss}

In this work, we treat the spin bath of P1 centers in a mean-field manner,
approximating it as a classical noise field acting on the NV center spin.
This treatment is justified because the coupling between the bath spins is of long-range character, and its magnitude is of the same order as the coupling of a bath spin to the central spin. In this case, the back-action from the central spin on the bath is negligible: the action of a single NV center on a given P1 spin is small in comparison with the action of hundreds of other bath spins. Next, due to the long-range character of the dipolar coupling, the NV spin experiences the action of a large number of P1 bath spins. As a result, the action of the bath as a whole on the NV spin can be approximated as a random magnetic field, which is Gaussian (due to the large number of bath spins acting on the NV spin with comparable strength), has zero mean, and whose variance is\cite{ZhangJPCM07} $b^2=(1/4)\sum_{k}a^2_k$. Due to small back-action, this random field is stationary. Moreover, since each bath spin is coupled to a large number of other bath spins, and the dynamics of different bath spins are incoherent, we expect this random magnetic field to be Markovian, i.e., history-independent.

This treatment follows the lines of the early works in the area of magnetic resonance.\cite{KlauderAnderson,Kubo1,HuHartmann} It has been numerically and experimentally justified for the system under consideration.\cite{Dobrovitski08,Hanson08,Dobrovitski09,deLange10}
We stress that the validity of such approximation
is built upon the assumption of small back-action from the central spin
and the lack of coherence between the dynamics of different bath spins.
If the couplings between the bath spins are much smaller
than the coupling between a single bath spin and the central spin,
then the bath dynamics is conditioned on the state of the central spin,
and the resulting correlations in the bath play a major role.\cite{RBLiu07,Witzel05,Cywinski09,Saikin07,Maze08b,CoishLoss}

The noise field $B(t)$, which is Gaussian, Markovian, and stationary, is represented by an Ornstein-Uhlenbeck (O-U) process\cite{vanKampen} with the correlation function
\begin{equation}
\label{eq:corr}
C(t)= \langle B(0)B(t)\rangle=b^2 \exp{(-|t|/\tau_c)}
\end{equation}
where $b$ describes the characteristic coupling strength of the central spin to
the bath, and $\tau_c$ is the correlation time of the noise field (governed by the coupling between the bath spins).
Thus, the Hamiltonian governing the dynamics of the central spin is
\begin{equation}
H=S^z_0 B_1(t) + H_c(t),
\end{equation}
where the term $H_c(t)$ represents the control pulses (specified below), and the field $B_1(t)$ includes the random bath field $B(t)$, the static applied field, and the hyperfine field $A_0 I_0^z$ from the NV's own nuclear spin.
Here and below we adopt the unit $\hbar=\gamma=1$,
where $\gamma$ is the gyromagnetic ratio for the NV center spin.

The performance of a DD sequence can be measured by
the overlap between the initial state and the state after the pulse sequence, $\Tr[\rho(0)\rho_S(t)]$,
where $\rho(0)=|\psi(0)\rangle\langle\psi(0)|$ is the initial density matrix of the central spin,
and $\rho_S(t)$ is the reduced density matrix of the central spin at time $t$;
we assume that the central spin is disentangled from the bath at $t=0$ and is prepared in the pure state $|\psi(0)\rangle$.
In our study, we focus on two initial states of the central spin,
with the spin directed along the $x$ and the $y$ axes, which we will denote
as $|X\rangle$ and $|Y\rangle$ respectively.
Any density matrix for a spin-1/2 system can be expressed as
\begin{equation}
\rho_S(t)=(1/2)\left[{\mathbf 1}+ S_X \sigma_x
+ S_Y \sigma_y+ S_Z \sigma_z\right],
\end{equation}
where $\mathbf 1$ is the identity operator, $\sigma_{x,y,z}$ are the Pauli operators, and the parameters $S_{X,Y,Z}$ determine the corresponding spin projections. Below, we use these parameters to characterize the decoupling fidelity: they are directly related to the standard fidelity measure,\cite{NielsenChuang} which in our setting acquires the form $F=\sqrt{\langle\psi(0)|\rho_S(t)|\psi(0)\rangle}$. The fidelities $S_X$ and $S_Y$ are normalized to the range $[-1,1]$, are directly determined experimentally, and coincide with $2F^2-1$ for the corresponding initial states.

Our analysis and numerical simulations have been performed
in a rotating frame with a frequency close to
the center of the ESR line of the NV center spin.
In this rotating frame, the average field $\langle B_1(t)\rangle$ is $B_{\tt dtn}+A_0I^z_0$, where $B_{\tt dtn}$ is the small detuning of the pulse field from the exact resonance. Its value $B_{\tt dtn}=-2\pi\cdot 0.5$~MHz has been obtained by fitting the experimental results.
In simulations, for different realizations of $B(t)$, we randomly sample the values of $I^z_0=+1$, $-1$, $0$ with the probabilities $p_+=0.5$, $p_-=0.2$, $p_0=0.3$, respectively. These values are determined experimentally, from the Ramsey fringe experiments.

\section{experimental setup}
\label{sec:exp}
We study NV centers in a Ib bulk diamond sample (Element Six) which is mounted in a scanning confocal microscope at room temperature.\cite{Jelezko04} The local density of electronic spins surrounding the NV center used in these experiments is $\sim 100$~ppm, which is estimated from the decay of Ramsey fringes of nitrogen spins. \cite{GijsBathCtrl} A static magnetic field ($B_0 = $114~G) is applied along the NV symmetry axis by a custom made vector magnet (Alpha Magnetics). 

The envelopes of the pulse sequences are generated using an arbitrary waveform generator (Tektronix AWG 5014B). The AWG is connected to the I/Q modulation inputs of a vector signal generator (R\&S SMBV100A) which has a 250 MHz modulation bandwidth. In order to achieve high Rabi frequencies we amplify the microwave signal using a high power amplifier (AR 25S1G4). The amplified signal is delivered to the sample by a coplanar waveguide (CPW) which is fabricated directly on the surface of the diamond.\cite{Fuchs09,deLange10} This configuration allows us to generate sequences with short (8 ns) $\pi$-pulses with arbitrary but well defined relative phases.


\section{Dynamical decoupling with ideal pulses}
\label{sec:scaling}

Analysis of DD is often based on the Magnus expansion (ME),\cite{Haeberlen}
which is a cumulant expansion of the evolution
operator of the system.
However, the Magnus expansion is not always sufficient for the analysis of DD under realistic conditions. As an asymptotic expansion, the ME is valid only for the sequences with very short period, satisfying the condition $T_c ||H_B||\ll 1$, where $T_c$ is the DD period, and $||H_B||$ is the norm of the bath Hamiltonian. In many realistic situations, $||H_B||$ is macroscopically large, and the formal ME validity condition requires unreasonably small $T_c$; nevertheless, DD performs very well far beyond the formal ME limits. In this work, by virtue of the known properties of the O-U process, we perform analytical and numerical study of different DD sequences without invoking ME.

Specifically, we study the following protocols: PDD XY (in which the central spin is rotated about the $x$ and $y$ axes alternately),
its symmetrized version SDD XY and concatenated CDD versions;
CPMG, XY4 (a DD scheme with the same timing as CPMG but based on the two-axis control),\cite{Gullion90}
its symmetrized version XY8 and its concatenated version CDD$^{\tt XY4}$. We also consider the aperiodic UDD sequence and its extension QDD.\cite{West10}

In this section, we assume the pulses in the DD sequences are ideal, i.e.\ without pulse errors,
and focus on the performance of DD sequences as a function of the inter-pulse delay.
Note that for ideal pulses, in the regime of pure dephasing,
the axes of the $\pi$-pulses in the DD sequence are irrelevant, as explained below. Thus, our results are equally applicable to the case of the single-axis control sequences (single-axis PDD, SDD, CDD, etc.) However, as we will see in the subsequent sections, the single-axis protocols perform poorly in the case of non-ideal pulses, so these protocols are of secondary importance for our purposes.

\subsection{Analytical expression for decoherence}
\label{sec:analy_general}

The dephasing dynamics of a spin 1/2 subjected to a classical O-U noise field $B(t)$ can be studied analytically. The transverse spin component precesses around the $z$-axis, and is rotated by an angle $\Phi(T)=\int_0^T B(t) dt$ at time $t=T$, so that
the spin components along $x$ and $y$ are determined by $\langle\cos{\Phi(T)}\rangle$ and
$\langle\sin{\Phi(T)}\rangle$, where the angular brackets denote the averaging over all realizations of the O-U process. Note that in this section, the static detuning $B_{\tt dtn}$ and the static hyperfine field are neglected: for perfect DD pulses, these static fields are completely removed by a single-pulse spin echo, and by any DD sequence.

For a O-U process with zero mean, $\langle\sin{\Phi}\rangle=0$, so that the decoherence leads only to the decay of the initial spin component, without extra rotation. I.e., if the central spin was prepared at $t=0$ along the $x$-axis
($S_X=1$, and $S_Y=S_Z=0$), then $S_Y=S_Z=0$ at all times, while
\begin{equation}
S_X(T) = F(T) = \left\langle\exp{\left(-i\int_0^T B(t)dt\right)}\right\rangle.
\end{equation}
The decay function $F(T)$ in the case of free evolution (no DD pulses) can be easily calculated:\cite{KlauderAnderson,Kubo1} the transition probability for a O-U process has Gaussian form, so the functional in the equation above is a standard Gaussian path integral. For a slowly fluctuating bath, with $b\tau_c\gg 1$,
the free decay of the central spin has Gaussian form $F(T)=\exp{(-b^2 T^2/2)}$.

In a spin echo experiment, after free evolution during time $\tau$,
a $\pi$ pulse is applied, flipping the sign of the $z$-component of the central spin, and the sign of the phase accumulation is inverted. The resulting change in the dynamics can be described by correspondingly rotating the coordinate frame and inverting the sign of the field $B(t)$.\cite{Slichter,Haeberlen} Thus, for a single-pulse Hahn echo experiment, of total duration $T=2\tau$, the decay of the transversal component of the central spin is determined by the functional \cite{KlauderAnderson}
$E(2\tau)=\left\langle\exp{\left(-i\int_0^\tau B(t)dt+i\int_\tau^{2\tau} B(t)dt\right)}\right\rangle$. This is another Gaussian path integral which can be calculated explicitly, and for a slow bath ($b\tau_c\gg 1$) has a form $E(T=2\tau)=\exp{(-T/T_2)^3}$, where the echo decay time
\begin{eqnarray}
\label{eq:T2}
T_2=\Big(\frac{12\tau_c}{b^2}\Big)^{1/3}.
\end{eqnarray}
Our experiments on the Ramsey fringes decay (measuring the free dephasing rate) and on the Hahn spin echo decay demonstrate the theoretically predicted behavior. The measurements of the free decay time ($T^*_2=0.3~\mu$s) and the echo decay time ($T_2=2.8~\mu$s) allow to determine the bath parameters:
\begin{eqnarray}
\label{eq:bRfitting}
\tau_c &=& 25~\mu\text s \nonumber\\
b &=& 3.3~\mu\text s^{-1} \;.
\end{eqnarray}

When the central spin is subjected to a DD sequence,
multiple $\pi$-pulses are applied sequentially, and each pulse inverts
the sign of $S^z_0$. As for the spin echo, the spin evolution is conveniently described in the toggling coordinate frame, which is rotated along with each spin rotation; correspondingly, the sign of the field $B(t)$ is inverted with each pulse.\cite{Haeberlen} Transforming the spin evolution in the toggling frame back to the standard rotating frame is not difficult: for ideal pulse it is enough to note that the two coordinate frames coincide at the end of every DD period. The spin evolution in this toggling coordinate frame remains essentially the same: the spin rotates in the $x$-$y$ plane, and the total rotation angle is $\tilde\Phi=\int_0^T \xi(t) B(t)dt$, where
$\xi(t)$ is the time-domain filter function, which takes into account the impact of the pulses: $\xi(0)=+1$, and changes sign every time when a $\pi$-pulse is applied to the central spin.

Again, for a O-U process with zero mean, the decoherence in presence of the DD pulses is reduced to the decay of the transverse spin component, without extra rotation. Considering the evolution of the central spin during the time interval $[0,T]$, the decoupling fidelity (equal to the transverse spin component) is
\begin{equation}
\label{eq:ST}
S(T) = \left\langle \exp{\big(-i\int_0^T \xi(s) B(s)ds\big)}
  \right\rangle,
\end{equation}
where we omitted the index of the transverse component as irrelevant.
For an O-U process $B(t)$, the averaging in Eq.~(\ref{eq:ST})
can then be carried out explicitly:\cite{Kubo1} this is the characteristic functional of a O-U process, which can be re-written in the form
\begin{equation}
S(T)=\exp{[-b^2\int_0^T {\rm e}^{-Rs} p(s)ds]}=\exp{[-b^2 W(T)]},
\end{equation}
where $R\equiv 1/\tau_c$, and the convolution integral
$p(s) = \int_0^{T-s} \xi(t)\xi(t+s) dt$
contains all relevant information about the pulse sequence.

We consider a decoupling sequence containing $N_c$ cycles,
each of duration $T_c$, so that $T=N_c T_c$. In this case, it is convenient to define a filter function $\xi_0(t)$ for a single period: this function is zero at $t<0$,
equal to +1 at $t=0$, changes sign every time a pulse is applied, and becomes zero at $t>T_c$. Then, the total function $\xi(t)$ for the $k$-th cycle equals to $\xi_0\big(t-(k-1)T_c\big)$, and $\xi_0(t)$ therefore fully characterizes the pulse sequence.
By calculating $p(s)$ in each cycle,
and taking advantage of the periodicity of the sequence,
the fidelity decay can be derived as
\begin{equation}
\label{eq:generalST}
W(T)=\Gamma_N (Q_{11}+Q_{12})- P_N Q_{12},
\end{equation}
where $\Gamma_N$ and $P_N$ are $N_c$-dependent constants given in Eq.~(\ref{eq:PN}) in Appendix~\ref{app:analysis}
and the integrals
\begin{eqnarray}
\label{eq:Q11}
Q_{11} &=& \int_0^{T_c} {\rm e}^{-Rs} q_{11}(s) ds \\ \nonumber
Q_{12} &=& \int_0^{T_c} {\rm e}^{-Rs} q_{11}(T_c-s) ds.
\end{eqnarray}
with
$
q_{11}(s) = \int_0^{T_c-s} \xi_0(t)\xi_0(t+s) dt
$.
See Appendix~\ref{app:analysis} for the details of the derivation.

For any DD sequence with a periodical structure,
it is then straightforward to calculate the decay of the decoupling fidelity,
Eq.~(\ref{eq:generalST}), starting from the specific filter function, $\xi_0(t)$.

Note that in the above derivation,
since we are dealing with pure dephasing and ideal pulses,
the axes of the $\pi$-pulses in the DD sequence
do not show up in the filter function.
Therefore a $\pi$-pulse about the $x$ axis (denoted as $\pi_\text X$)
and that about the $y$ axis ($\pi_\text Y$)
result in the same decoupling performance.
It is the timing of the pulses that matters.

\subsection{DD  protocols based on equidistant pulses}
\label{sec:pddxy}
We now study the performance of PDD XY,
its concatenated version CDD, and its symmetrized version, SDD XY.
While the results here are equally applicable to the corresponding single-axis protocols, the two-axis control is the main focus of our study.

Within the PDD XY protocol, the sequence of pulses
d-$\pi_\text X$-d-$\pi_\text Y$-d-$\pi_\text X$-d-$\pi_\text Y$
is repeated many times, where d denotes a free evolution of duration $\tau$.
Following the procedure in Sec.~\ref{sec:analy_general},
we can derive the dephasing exponent $W_{\tt PDD}(t)$.
In our experiments, the typical inter-pulse delay is $\sim 0.02$--$1~\mu$s,
which is much shorter than the bath correlation time, $\tau_c=25~\mu$s.
We thus focus on the case $R\tau\ll 1$.
When the number of cycles $N_c$ is small, $RT=N_cRT_c\ll 1$ (we will refer to this regime as short times), the decoupling fidelity decays as
\begin{equation}
\label{eq:SPDD}
S(T)=\exp\Big[-\frac{4}{N_d^2}\Big(\frac{T}{T_{2}}\Big)^3\Big]
\end{equation}
where $T_2$ is the spin echo decay time, Eq.~(\ref{eq:T2}),
and $N_d$ is the total number of the inter-pulse delays (equal to the total number of pulses).
When $N_c$ becomes so large that $RT=RN_cT_c\gg 1$,
(we refer to this regime as long times),
the decay rate becomes four times smaller:
\begin{equation}
\label{eq:SPDD1}
S(T)=\exp\Big[-\frac{1}{N_d^2}\Big(\frac{T}{T_{2}}\Big)^3\Big]
\end{equation}
Thus, for PDD XY, as the number of period increases,
the decay slows down (detailed derivation is in Appendix~\ref{app:XY}).
This feature is shown in Fig.~\ref{fig:PDDScaling}, where numerical results for $S(t)$
are shown as a function of $N_d$, with fixed inter-pulse delay $\tau$.

\begin{figure}[htbp]
\includegraphics[height=4cm, angle=0]{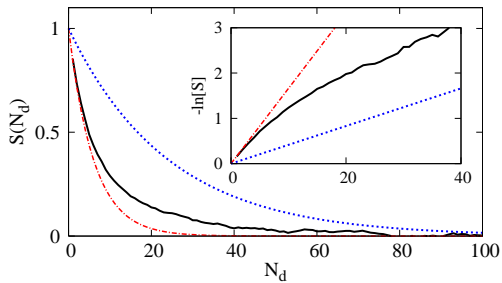}
\caption{\label{fig:PDDScaling}
(Color online).
Simulation results for the decoupling fidelity as a function
of the number $N_d$ of the inter-pulse delays in PDD XY (solid black line).
The dashed red line corresponds to the analytical expression (\ref{eq:SPDD}) for small $N_d$, and the dotted blue line corresponds to the expression (\ref{eq:SPDD1}) for large $N_d$.
The duration of the inter-pulse delay is $\tau=0.6~\mu$s.
The inset shows the same curves with logarithmic scale in the vertical axis,  demonstrating that the solid black line (simulation results) is parallel to the dotted blue line (analytics for long times) after $N_d\sim 20$, i.e.\ the decoherence rate corresponds to the long-time regime, Eq.~(\ref{eq:SPDD1}).
}
\end{figure}

We now consider concatenated protocols based on PDD XY.
The concatenated sequence of level $\ell$, which we denote as CDD$_{\ell}$,
is constructed recursively as\cite{Khodjasteh05,Khodjasteh07}
\begin{equation}
\text{(CDD$_{\ell-1}$)-$\pi_\text X$-(CDD$_{\ell-1}$)-$\pi_\text Y$-(CDD$_{\ell-1}$)-$\pi_\text X$-(CDD$_{\ell-1}$)-$\pi_\text Y$}\;,
\end{equation}
starting from PDD XY as the first-level CDD sequence.
Extention to larger times can be achieved either by increasing the concatenation level, or by periodically repeating CDD of a fixed level, since the length of the CDD period $T_c$ increases exponentially with $\ell$.
The detailed analysis of this family of protocols is given in Appendix~\ref{app:XY}: it is performed recursively, and
the decay rate for CDD$_\ell$ can be calculated from that of CDD$_{\ell-1}$.
In the most experimentally relevant case, when $RT_c\ll 1$, we find that for all concatenation levels,
the decay rate for both short times and long times are the same, equal to the
short-time decay rate of PDD XY, Eq.~(\ref{eq:SPDD}).
This behavior is exactly the opposite of a general expectation based on ME, that by increasing the CDD order, the decoupling performance should improve. The origin of such discrepancy is presented later in Sec~\ref{sec:xy4}.

\begin{figure}[htbp]
\includegraphics[height=4cm, angle=0]{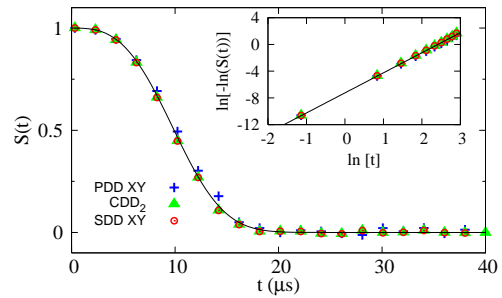}
\caption{\label{fig:XYPDD}
(Color online).
Simulation results for the decoupling fidelity as a functions of the total evolution time
for PDD XY (blue crosses), CDD$_2$ (green triangles), and SDD XY (red circles). In all protocols $N_d=16$, and $\tau$ is increasing with $t$. The two latter sequences produce identical results.
The black line shows the analytical results, Eq.~(\ref{eq:SPDD}).
Inset shows the same results in log-log scale.
}
\end{figure}

The symmetrized protocol, SDD XY, is constructed by taking the PDD XY period, and adding to it the time-inverted PDD XY period, producing the sequence
with the period
\begin{eqnarray}
\label{eq:SDDXYseq}
\text {(d-$\pi_\text X$-d-$\pi_\text Y$-d-$\pi_\text X$-d-$\pi_\text Y$)-
($\pi_\text Y$-d-$\pi_\text X$-d-$\pi_\text Y$-d-$\pi_\text X$-d)}\;,
\end{eqnarray}
which has exactly the same timing of the pulses as the half-period of CDD$_2$
(since for ideal pulses $\pi_\text Y\pi_\text Y=\mathbf 1$).
The filter function for SDD XY therefore coincides with the filter function of  CDD$_2$, and so is the decoupling fidelity.
Fig.~\ref{fig:XYPDD} shows the numerical results
for the performance of PDD XY, CDD$_2$ and SDD XY, each with $N_d=16$.
All curves agree with the analytical expression Eq.~(\ref{eq:SPDD}).

\subsection{Sequences based on CPMG timing}
\label{sec:xy4}
\begin{figure}[htbp]
\includegraphics[height=8.5cm, angle=270]{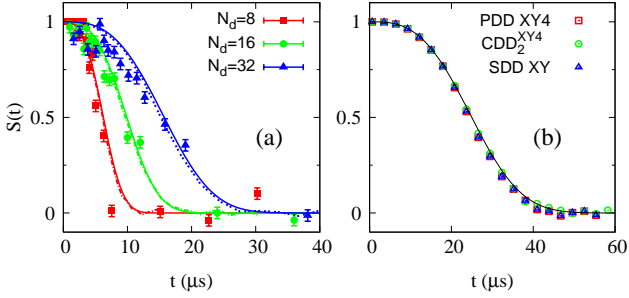}
\caption{\label{fig:XY4}
(Color online).
Decoupling fidelity as a functions of the total evolution time.
(a) XY4 for $N_d=8,~16$ and $32$ (red, green, and blue lines and symbols, correspondingly).
Solid lines are analytical results obtained from Eq.~(\ref{eq:scaling}),
the dashed lines are numerical results,
and the symbols are the experimental data.
(b) Numerical results for XY4 with 8 periods (red squares),
CDD$_2^{\tt XY4}$ (green circles), and XY8 (blue triangles) protocols with 4 periods. The number of delays is $64$ for all protocols.
The black line shows the analytical prediction of Eq.~(\ref{eq:scaling}). As predicted, all three sequences demonstrate the same performance.
}
\end{figure}

The period of CPMG sequence is d-$\pi$-d-d-$\pi$-d.
The sequence XY4 has the same pulse timing,
but is based on the two-axis control:\cite{Gullion90}
\begin{equation}
\label{eq:pddxy4}
\text{d-$\pi_\text X$-d-d-$\pi_\text Y$-d-d-$\pi_\text X$-d-d-$\pi_\text Y$-d}\;.
\end{equation}
Since we are considering ideal pulses, and limited to dephasing of the central spin,
these two sequences are equivalent.
Detailed analysis (in Appendix~\ref{app:XY4}) shows that
the decay rate of the fidelity for both short and long times
is the same. If we denote $T_{1/e}$
as the $1/e$ decay time of the decoupling fidelity,
then for a sequence with CPMG timing, the fidelity decays as
\begin{eqnarray}
\label{eq:scaling}
S(T)=\exp\Big[-\Big(\frac{T}{T_{1/e}}\Big)^3\Big]
\end{eqnarray}
with
\begin{eqnarray}
\label{eq:Te}
T_{1/e}=\left(\frac{N_d}{2}\right)^{2/3}T_2\;.
\end{eqnarray}
This is exactly the scaling behavior of the decay time observed in our experiments.
Fig.~\ref{fig:XY4}(a) shows the experimental data and the simulation results
for $N_d=8,~16,~32$, which are in excellent agreement with Eq.~(\ref{eq:scaling}).

Let us compare XY4 with PDD XY.
If we take the inter-pulse delays $\tau_{\tt PDD}=2\tau_{\tt XY4}$,
then the only difference
between the two sequences will be the very first and the
very last segments, each of duration $\tau_{\tt XY4}$.
Eq.~(\ref{eq:scaling}) is then equivalent to Eq.~(\ref{eq:SPDD1}),
taking into account that $2N_d^{\tt PDD}=N_d^{\tt XY4}$.
Therefore, at short times, the decay for PDD XY sequence is four times
faster than for XY4. Note that this difference is the result of only two segments at the very beginning and at the very end of the sequence.
For long times (large $N_d$),
the decay rates of PDD XY and XY4 are the same, but the absolute value of the DD fidelity is exponentially higher for the CPMG-timed sequence XY4.

The symmetrized version of XY4 sequence, often referred to as  XY8,\cite{Gullion90}
has a period
\begin{eqnarray}
\label{eq:XY8seq}
\text {(d-$\pi_\text X$-d-d-$\pi_\text Y$-d-d-$\pi_\text X$-d-d-$\pi_\text Y$-d)-}\nonumber\\
\text{(d-$\pi_\text Y$-d-d-$\pi_\text X$-d-d-$\pi_\text Y$-d-d-$\pi_\text X$-d)}\;.
\end{eqnarray}
Since its filter function is the same as XY4, the fidelity decay is also the same. For concatenated version of XY4, the short-time and long-time decay rates
in CDD$^{\tt  XY4}_{\ell}$ for any level are the same,
equal to that in XY4 (See Appendix~\ref{app:XY4}).
Fig.~\ref{fig:XY4}(b) shows the simulation results for XY4,
XY8, and CDD$^{\tt  XY4}_{2}$. The number of delays in all sequences
is the same: $N_d=64$ and the results are in agreeement with Eq.~(\ref{eq:scaling}).

Therefore, we arrive at the conclusion that
for the spin bath considered in our study (Gaussian Markovian noise field with the Lorentzian spectrum), the performance of both the PDD- and the CPMG-based sequences is not improved neither by symmetrization or concatenation, and
the simple XY4 sequence provides the optimal choice for experiments.
This result may look somewhat discouraging, and is in
contrast with the standard qualitative expectation, based on the Magnus expansion, that the higher-order sequences should generally provide better fidelity. It is also in clear contrast with the ME prediction that the decay exponent $W(T)$ for symmetrized sequences should contain only even-order terms, while our calculations produce $W(T)\sim T^3$. The problem here is the inapplicability of the Magnus expansion. The spectral density of our spin bath has a Lorentzian shape, $\sim 1/(R^2 + \omega^2)$, which has formally infinite second moment. In terms of Magnus expansion, this corresponds to a formally infinite $||H_B||$. In reality, of course, the power spectrum of the spin bath is limited by the fastest possible flip-flop rate between two bath spins, which is achieved when the two P1 centers are located at the nearest sites in the diamond lattice, but this frequency is in the GHz range, and is far above all other timescales relevant for decoherence. This situation is in sharp contrast with, e.g.\ decoherence of the electron spin by a bath of nuclear spins in quantum dots.\cite{Witzel05,RBLiu07,WYang08,Cywinski09,Cywinski10,Cywinski11,CoishBaugh,CoishLoss}

\subsection{Aperiodic decoupling sequences}
In Uhrig DD scheme of level $\ell$, denoted as UDD$_\ell$,
a sequence of $\pi$ pulses rotate the spin about the same axis (taken here as $x$), and the pulses are applied at times
\begin{eqnarray}
\label{eq:UDDtiming}
t_j=T\sin^2\Big(\frac{j\pi}{2N_p+2}\Big),
\end{eqnarray}
where $T$ is the total evolution time, and
$j=1,2,\ldots,\ell$ for even $\ell$ and $j=1,2,\ldots,\ell+1$ for odd $\ell$.
The generalization of the UDD for the case of the two-axis control,
the QDD sequence,\cite{West10} is composed of two nested UDD sequences,
during which the central spin is rotated about two perpendicular axes respectively; to be concrete, here we consider
the QDD sequence with $\pi_\text X$-pulses in the outer hierarchical level,
and $\pi_\text Y$-pulses in the inner hierarchical level. Thus,
for odd level $\ell=2n-1$,
the QDD$_\ell$ sequence is
\begin{equation}
\label{eq:seq_QDDodd}
\text{UDD}_{\ell}^{(\text Y)}(\tau_1)\text -\pi_\text X\text-\text{UDD}_{\ell}^{(\text Y)}(\tau_2)\text -\pi_\text X\cdots\text-\text{UDD}_{\ell}^{(\text Y)}(\tau_{\ell+1})\text -\pi_\text X
\end{equation}
where $\sum_{j=1}^{\ell+1}\tau_j=T$ and
$\text{UDD}_{\ell}^{(\text Y)}(\tau_j)$ denotes UDD$_\ell$ based on $\pi_\text Y$ pulses with evolution time $\tau_j$.
The division of the total time $T$ into intervals $\tau_j$ satisfies the same rule as for UDD,
i.e., $\tau_j=t_j-t_{j-1}$, with $t_j$ given by Eq.~(\ref{eq:UDDtiming}).
The number of pulses in such a QDD sequence is $N_p=(\ell+1)(\ell+2)$.
For even level $\ell=2n$, the QDD$_\ell$ sequence is
\begin{equation}
\label{eq:seq_QDDeven}
\text{UDD}_{\ell}^{(\text Y)}(\tau_1)\text -\pi_\text X\text-\cdots\text-\text{UDD}_{\ell}^{(\text Y)}(\tau_{\ell})\text -\pi_\text X \text- \text{UDD}_{\ell}^{(\text Y)}(\tau_{\ell+1})
\end{equation}
and the number of pulses is $N_p=\ell(\ell+2)$.

\begin{figure}[htbp]
\includegraphics[height=8.5cm, angle=270]{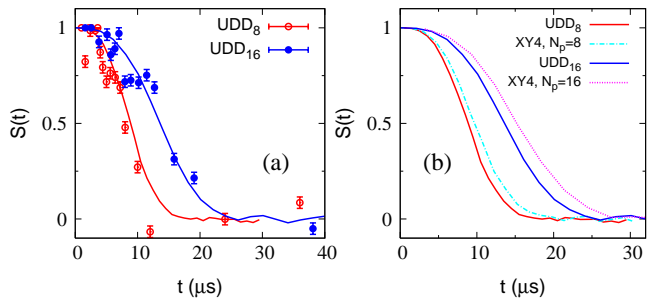}
\caption{\label{fig:UDDvsXY4}
(Color online).
(a) Decoupling fidelity as a functions of the total evolution time
for UDD with $N_p=8$ and $N_p=16$ pulses (red and blue, respectively). Dots are the experimental data,
and lines are simulation results.
(b) Comparison of the simulation results for UDD (solid lines)
and XY4 sequence (dashed lines) with the same number of pulses (red for $N_p=8$ and blue for $N_p=16$).
}
\end{figure}

\begin{figure}[htbp]
\includegraphics[height=8.5cm, angle=270]{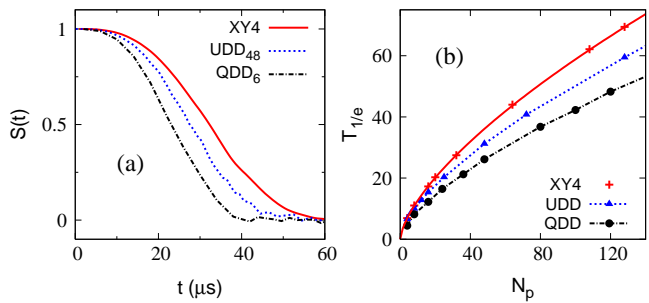}
\caption{\label{fig:QDDvsUDD}
(Color online).
Simulation results. (a) Decoupling fidelity as a functions of the total evolution time
for XY4 with 12 periods, UDD$_{48}$ and QDD$_6$. Each sequence has $N_p=48$ pulses.
(b) The decay time $T_{1/e}$, as a function of the number of pulses $N_p$ for XY4 (solid line), UDD (dotted line), and QDD (dash-dotted line).
Solid line shows the scaling relation given by Eq.~(\ref{eq:Te}), with $T_2$ fitted to be $2.73~\mu$s.
}
\end{figure}

The analysis in Sec.~\ref{sec:analy_general}
is not applicable to UDD and QDD due to the lack of periodicity in these sequences.
We perform only numerical study on these two sequences.
Fig.~\ref{fig:UDDvsXY4} shows the decoupling performance
of UDD, in comparison with XY4 with the same number of pulses, $N_p=8,~16$.
For both cases, the fidelity decay in UDD is faster than in XY4.
The fidelity decay in QDD is even faster than UDD,
given that the number of pulses is the same, see Fig.~\ref{fig:QDDvsUDD}(a).
The $1/e$ decay time of UDD, XY4 (CPMG), and QDD are shown in Fig.~\ref{fig:QDDvsUDD}(b).
For all number of pulses we have studied, the XY4 sequence performs better than UDD and QDD. Since the noise field $B(t)$ generated by O-U process
has Lorentzian spectral density without sharp cutoff,
our results agrees with the existing  knowledge,\cite{Pasini10,Cywinski08,Cywinski09,WYang08,Biercuk09PRA,Uys09,deLange10} that for the bath spectrum with a soft cutoff, the
pulse sequences with CPMG timing outperform UDD.

\section {Effects of the pulse error accumulation}
\label{sec:err}
In this section, we focus on the effect of the pulse error accumulation for the DD sequences introduced above.

\subsection{Model for pulse error description}
\label{sec:errAnaly}

In our analysis, we assume that the control pulses have infinitely small duration, and therefore treat them as rotation operators.
The imperfections in the pulses are taken into account
by considering the errors in the rotation axis and rotation angles.\cite{Tyryshkin10,Wang10,Dobrovitski10}
For instance, for a nominal $\pi$-pulse about the $x$ axis (the $\pi_\text X$ pulse), the actual evolution of the central spin is expressed as
\begin{eqnarray}
\label{eq:piX}
U_{\text X}&=&\exp{[-i(\pi + \epsilon_x)({\bf S}\cdot\vec {\bf n})]}\\ \nonumber
\end{eqnarray}
where $\vec{\bf n}=(\sqrt{1-n_y^2-n_z^2},n_y,n_z)$
is the rotation axis which slightly deviates from the $x$ axis ($n_y$ and $n_z$ are small),
and $\epsilon_x$ is the error in the rotation angle.
Similarly, the rotation operator for a nominally $\pi$-pulse about the $y$ axis (the $\pi_\text Y$ pulse) is
\begin{eqnarray}
\label{eq:piY}
U_{\text Y}&=&\exp{[-i(\pi + \epsilon_y)({\bf S}\cdot\vec {\bf m})]}\;,
\end{eqnarray}
where $\epsilon_y$ is the rotation angle error
and $\vec{\bf m}=(m_x,\sqrt{1-m_x^2-m_z^2},m_z)$ is the actual rotation axis.

In the studies below, the pulse errors $\epsilon_x$, $\epsilon_y$, $n_y,~m_x$ are kept constant for all experimental runs ($\sim 10^6$ in our experiments),
and during each single run. This corresponds to an excellent stability of the pulse generating hardware used in our experiments. However, the components $n_z$ and $m_z$ or the rotation axes are noticeably affected by the hyperfine coupling $A_0S^zI^z_0$. Correspondingly, these two errors are treated as static during each run,
but have different values for different runs,
depending on the values of $I^z_0=1,~0,~-1$ as
$n_z=n_0I^z_0$ and $m_z=m_0I^z_0$.
The parameters taken in the simulation are
$\epsilon_x=\epsilon_y=-0.02$, $m_x=0.005$, $n_0=m_0=0.05$, and $n_y=0$, being determined from the bootstrap protocol for pulse characterization.\cite{Dobrovitski10}

Note that for imperfect pulses, it becomes important how the $\pi$ rotations are implemented, via $\pi_X$ or via $\pi_Y$ pulses. Therefore, in contrast with the previous section, the sequences with the single-axis and with the two-axis control will be considered separately.

\subsection{CPMG sequence}
\label{sec:ErrCPMG}

The evolution operator for a period of the CPMG sequence is
\begin{eqnarray}
U^{\tt CPMG}=  U_{\text d}(\tau)U_{\text X} U_{\text d}(\tau) U_{\text d}(\tau)U_{\text X} U_{\text d}(\tau)\;.
\end{eqnarray}
To gain qualitative insight into the problem of the pulse error accumulation, let us assume that the noise field $B$ is static.
Keeping only the terms of the first order in the pulse errors,
\begin{eqnarray}
\label{eq:CPMG}
U^{\tt CPMG}=-{\mathbf 1}+i(\epsilon_x\cos{\phi_{\tt d}}+2n_z\sin{\phi_{\tt d}})\sigma^x\;,
\end{eqnarray}
where $\phi_{\tt d}=B\tau$ is the noise phase accumulated during an inter-pulse delay.
For perfect pulses, the evolution operator is identity, and the quantum state of the central spin is preserved perfectly for static noise.
For imperfect $\pi_{\text X}$ pulses, the evolution during a CPMG period corresponds to a spin rotation about the $x$ axis
by angle $2\pi+\theta_{\tt CPMG}$ with
$\theta_{\tt CPMG}=-2(\epsilon_x\cos{\phi_{\tt d}}+2n_z\sin{\phi_{\tt d}})$.
Repeating $N_\text p$ periods, the overall evolution of the central spin is equivalent to a rotation about the same axis
by an angle $N_\text p\theta_{\tt CPMG}$.

To examine how different spin states are preserved in DD,
we study two perpendicular spin components $S_{\text X}$ and $S_{\text Y}$.
Since the $x$-component is not affected by rotations about $x$ axis,
the state $|X\rangle$ is preserved very well, being spoiled only by the higher-order terms.
For the initial state $|Y\rangle$, in a single realization of B,
the $y$ component is
\begin{equation}
S_{Y,{\tt single}} =\cos\theta_{\tt CPMG}\;,
\end{equation}
i.e, the decoupling fidelity oscillates at a frequency determined by the pulse errors. After averaging $S_{Y,{\tt single}}$ over different realizations of $B$, the fidelity decays due to the pulse error accumulation, as known from the early days of magnetic resonance.\cite{Slichter}
I.e., the spin component $S_{\text X}$ is well preserved in a CPMG sequence,
while $S_{\text Y}$ (when it becomes the CP sequence)
is sensitive to the pulse errors $\epsilon_x$ and $n_z$.

\begin{figure}[htbp]
\includegraphics[height=4cm, angle=0]{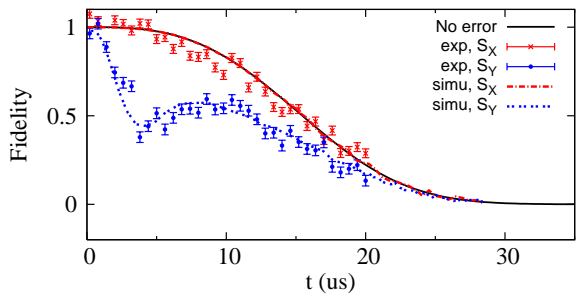}
\caption{\label{fig:CPMG}
(Color online).
Decoupling fidelity of CPMG sequence as a function of the total evolution time
for initial states $|X\rangle$ (red symbols and lines) and $|Y\rangle$ (blue symbols and lines). The number of pulses is $N_p=16$. Broken lines are the simulation results, the dots are the experimental results. The black solid line shows the results for ideal pulses.
}
\end{figure}

To study the experimental case of the dynamical noise (whose parameters are given by Eq.~\ref{eq:bRfitting}) in presence of the pulse errors, we performed numerical simulations. The results are shown in Fig.~\ref{fig:CPMG} for $N_p=16$ pulses applied to the states $|X\rangle$ and $|Y\rangle$. The total evolution time is varying from 0 to $15~\mu$s. The simulations agree
with the experimental results.

\subsection{Sequences based on PDD XY and XY4}

Analysis of the pulse error accumulation for PDD XY sequence, in the case of static noise, has been carried out in Ref.~[\onlinecite{Tyryshkin10}], and here we only briefly remind the main results.
The evolution operator for a single period,
to the first order in the pulse errors, is
\begin{eqnarray}
\label{eq:XY}
U^{\tt XY}&=& -{\mathbf 1}-2i(m_x+n_y)\sigma^z\;,
\end{eqnarray}
corresponding to a spin rotation about the $z$ axis
by the angle $2\pi+4(m_x+n_y)$. Thus, only
the in-plane components of the rotation axis $n_y$ and $m_x$
accumulate in the course of decoupling. In experiments, these errors can be minimized using appropriate pulse tuning.\cite{BurumEtal81,ShakaEtal88,Khaneja05,MortonSPAM,Dobrovitski10,FortunatoCoryPulses} Moreover, in Eq.~(\ref{eq:XY}) the evolution operators are independent of $\phi_d$, thus being insensitive to the resonance offset, the value of $B$, and the duration of the inter-pulse delay, $\tau$.

For the XY4 sequence, in spite of different timing, the pulse errors accumulate  in the same way (in the first order in the pulse errors), and the evolution operator for XY4 is also given by Eq.~(\ref{eq:XY}) in the case of static noise.
Therefore, XY4 is also very robust with respect to most pulse errors. In addition, due to better pulse timing, it demonstrates very good performance in the case of dynamic noise, as shown in Sec~\ref{sec:scaling}. Thus, we expect that this sequence is close to optimal for experiments: it combines simplicity, good scaling properties, and robustness to the pulse imperfections.

\begin{figure}[htbp]
\includegraphics[height=8cm, angle=270]{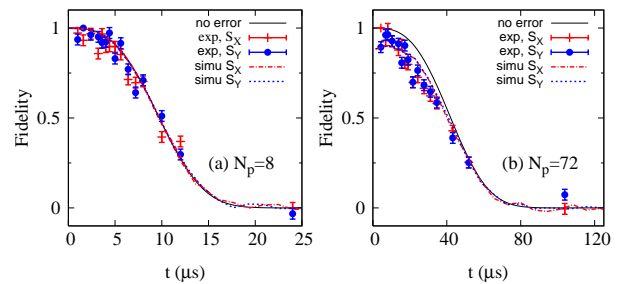}
\caption{\label{fig:PDDXY4}
(Color online).
Decoupling fidelity of the XY4 sequence as a function of the total evolution time for the initial states $|X\rangle$ (red symbols and lines) and $|Y\rangle$ (blue symbols and lines), in the presence of the pulse errors. The broken lines are the simulation results, and the dots are the experimental data. The black solid line shows the results for ideal pulses. (a) $N_p=8$. (b) $N_p=72$.
}
\end{figure}

Figure~\ref{fig:PDDXY4} shows the decoupling fidelity
as a function of the total evolution time for the XY4 sequence with imperfect pulses. The results for the initial states $|X\rangle$ and $|Y\rangle$ with $N_p=8$ and $N_p=72$ pulses are shown.
Since the in-plane pulse errors in our experiments are very small,
XY4 serves as a good decoupling sequence for the NV center spin
in a spin bath composed of P1 centers.

\begin{figure}[htbp]
\includegraphics[height=8.5cm, angle=270]{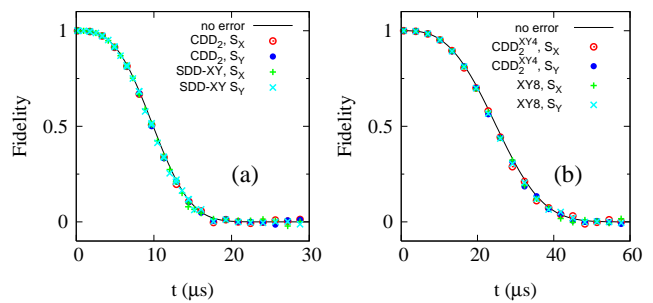}
\caption{\label{fig:CDDSDD}
(Color online).
Decoupling fidelity as a function of the total evolution time
for initial states $|X\rangle$ and $|Y\rangle$, in the presence of the pulse errors. The results are shown for SDD XY (2 periods) and CDD$_2$ sequences (a), and for XY8 (4 periods) and CDD$_2^{\tt XY4}$ sequences (b) with the number of pulses $N_p=16,~20,~32$ and $36$ respectively.
Black solid line shows the results for ideal pulses.
}
\end{figure}

The symmetrized version of PDD XY, SDD XY demonstrates worse scaling properties in the case of perfect pulses. However, it has advantages when the pulses are imperfect, as the first order terms in the pulse errors disappear in
the evolution operator. In the case of static noise field, to the second order,
\begin{eqnarray}
\label{eq:XYSDD}
U^{\tt SDDXY} &=& {\mathbf 1}+2i(m_x+n_y)\epsilon_y\sigma^x \\\nonumber
& &+2i(m_x+n_y)(\epsilon_x \cos\phi_{\tt d} + 2n_z\sin\phi_{\tt d})\sigma^y\;.
\end{eqnarray}
The evolution operator for the symmetrized version of XY4, the XY8 sequence, has even more symmetric form (again, for static noise):
\begin{eqnarray}
\label{eq:XY8}
U^{\tt XY8}&=& {\mathbf 1}+2i(m_x+n_y)(\epsilon_y \cos\phi_{\tt d}+2m_z\sin\phi_{\tt d})\sigma^x\nonumber\\
& &+2i(m_x+n_y)(\epsilon_x \cos\phi_{\tt d} + 2n_z\sin\phi_{\tt d})\sigma^y\;.
\end{eqnarray}
Thus, XY8 presents a good choice for the experimentalist when the pulse errors are not very small: this sequence is still very simple to program, it has the same good performance as XY4, but is sensitive to the pulse errors only in the second order.

The error accumulation in the concatenated version of PDD XY has
been analyzed before.\cite{Khodjasteh05,Khodjasteh07,Tyryshkin10}
An important feature is that CDD has an error-correction structure.
For the static noise, in the first order, the pulse errors do not accumulate
as the concatenation level increases, although the number of pulse increases exponentially. For the concatenated sequences based on XY4, we found that the same feature holds.
The evolution operator for CDD$^{\tt XY4}_\ell$
is still described by Eq.~(\ref{eq:XY}) for any $\ell$. While the leading terms are of the first order in the pulse errors, they do not accumulate as the concatenation order or the number of periods increase.

Fig.~\ref{fig:CDDSDD} shows the simulation results for the dynamic noise, with the parameters given by Eq.~\ref{eq:bRfitting}, for SDD XY (2 periods), CDD$_2$, XY8 (4 periods) and CDD$_2^{\tt XY4}$ with pulse number $N_p=16,~20,~32$ and $36$ respectively. Due to smallness of $m_x$ and $n_y$ in our simulations,
the pulse errors have little effect on the decoupling fidelity, and all sequences perform almost as if there was no pulse errors.

\begin{table*}[htbp]
\begin{threeparttable}
\caption{\label{tb:err}
The pulse errors influencing the DD fidelity in the first order for the initial states $|X\rangle$ and $|Y\rangle$, for the corresponding spin components $S_X$ and $S_Y$.
}
\begin{ruledtabular}
\begin{tabular}{ccccccccccc}
 & CPMG & XY,XY4 & $\text{UDD}^*$& (even $\ell$) &(odd $\ell$) & $\text{QDD}(T\to 0)^{**}$& (even $\ell$) & (odd $\ell$) & SDD,XY8 & CDD (XY, XY4) \\
 \hline
$S_X$ & 0 & $m_x$, $n_y$ & &0 &$\epsilon_x$, $n_z$ &&$\epsilon_x$, $\epsilon_y$ &0 & 0 & $m_x$, $n_y$\\
$S_Y$ & $\epsilon_x$, $n_z$ & $m_x$, $n_y$ && $\epsilon_x$, $n_z$ & $\epsilon_x$, $n_z$ && $\epsilon_x$, $\epsilon_y$& $\epsilon_x$ & 0 & $m_x$, $n_y$\\
\end{tabular}
\end{ruledtabular}
\begin{tablenotes}
  \item[*] UDD based on $\pi_\text X$-pulses
  \item[**] QDD with $\pi_\text X$-pulses in the outer hierarchical level, and $\pi_\text Y$-pulses in the inner hierarchical level.
 \end{tablenotes}
 \end{threeparttable}
\end{table*}

\subsection{UDD and QDD}

Since the inter-pulse delay in UDD is not uniform,
we use only the total evolution time $T$ in the following analysis.
As above, we present analytical results only for the case of static noise,
while the simulations show the results for the dynamical spin bath characterized by Eq.~\ref{eq:bRfitting}.

\begin{figure}[htbp]
\includegraphics[height=8.5cm,angle=270]{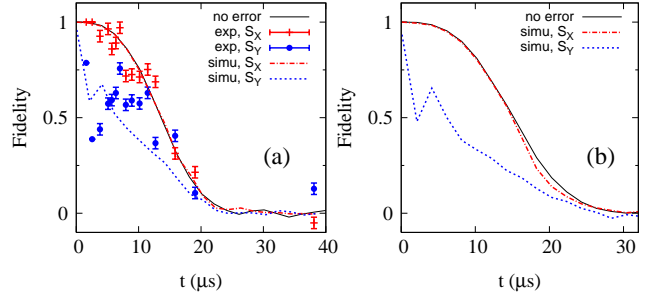}
\caption{\label{fig:UDD}
(Color online).
Decoupling fidelity for initial states $|X\rangle$ and $|Y\rangle$ in the presence of pulse errors.
Dots are crosses are the experimental data, and the broken and dotted lines are the simulation results for $S_X$ and $S_Y$, respectively.
The solid line shows the simulation results for perfect pulses. (a) UDD$_{16}$. (b) UDD$_{19}$.
}
\end{figure}

If the level of UDD is even, $\ell=2n$, then the pulse number is equal to $\ell$.
The evolution operator for a single electron spin, to first order in the pulse errors, is
\begin{equation}
\label{eq:UDDnEven}
U^{\tt UDD}_{2n}=(-1)^{n}\Big[{\mathbf 1}-i\sigma^x\theta_x \Big]\;,
\end{equation}
where $\theta_x$ is a linear combination of pulse errors $\epsilon_x$ and $n_z$,
with coefficients depending on $n$ and $BT$.
The spin is thus rotated about the $x$ axis by an angle $2\theta_x$,
which is dependent on $\epsilon_x$ and $n_z$.
Comparing to the CPMG sequence analyzed in Sec.~\ref{sec:ErrCPMG},
we see that for UDD with even number of the $\pi_\text X$-pulses, the
spin state $|X\rangle$ is well preserved
while the fidelity for $|Y\rangle$ state is strongly affected by the pulse error accumulation,
and the influencing factors are
the rotation angle error and the $z$-component of the rotation axis error.
Fig.~\ref{fig:UDD}(a) shows the decoupling performance of UDD$_{16}$ for the dynamical spin bath.

For $\ell=2n-1$, the pulse number $N_p$ equals $2n$
(there is a $\pi$ pulse at the end of the whole evolution).
The evolution operator is then
\begin{eqnarray}
\label{eq:UDDnOdd}
U^{\tt UDD}_{2n-1}&=& (-1)^{n}
({\mathbf 1}-i\theta_n\sigma^x-i\eta_n\sigma^y)
\end{eqnarray}
where
$\theta_n$ and $\eta_n$
are linear combinations of $\epsilon_x$ and $n_z$,
with coefficients depending on $n$ and $BT$.
This operator represents a rotation in the $x$-$y$ plane,
with the axis depending on the relative magnitudes of $\theta_n$ and $\eta_n$.
If $\theta_n\gg \eta_n$, then the axis is closer to the $x$ axis,
so that $S_X$ is preserved better than $S_Y$ (and vice versa).
We notice that in the limit $T\to 0$, $\eta_n$ vanishes, hence
the pulse error affect $S_X$ at larger times, while
$S_Y$ is affected already at small times,
see Fig.~\ref{fig:UDD}(b).

In the limit $T\to 0$, in which case the sequence is simply
successive pulses, the sequences with $\ell=2n-1$ and $\ell=2n$
are the same, and the evolution operator in the first order involves only $\epsilon_x$:
\begin{eqnarray}
\lim_{T\to 0} U^{\tt UDD}_{\ell}&=&(-1)^n \left( 1-in\epsilon_x \sigma^x \right)\;.
\end{eqnarray}
which represents a rotation about the $x$ axis by an angle $2n\epsilon_x$.
Hence for UDD sequences of large level, even for small total evolution time,
the accumulation of pulse errors in the rotation angle
would lead to fidelity loss for the spin state $|Y\rangle$.

\begin{figure}[htbp]
\includegraphics[height=8cm,angle=270]{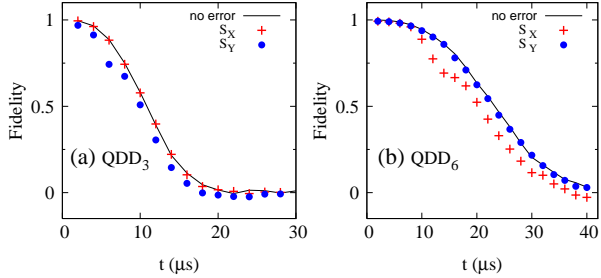}
\caption{\label{fig:QDD3QDD6}
(Color online).
Simulation results. Decoupling fidelity for the initial states $|X\rangle$ (red crosses) and $|Y\rangle$ (blue dots), in the presence of pulse errors.
Solid line is the results with ideal pulses.
(a) QDD$_3$.
(b) QDD$_6$.
}
\end{figure}

Now we consider QDD sequences
with $\pi_\text X$-pulses in the outer hierarchical level,
and $\pi_\text Y$-pulses in the inner hierarchical level
in the presence of the pulse errors.
The sequence of QDD$_\ell$ with $\ell=2n-1$ is shown in Eq.~(\ref{eq:seq_QDDodd})
with pulse number $N_p=(\ell+1)(\ell+2)$,
and for $\ell=2n$ in Eq.~(\ref{eq:seq_QDDeven})
with pulse number $N_p=\ell(\ell+2)$.

In the limit $T\to 0$, the evolution operators for QDD of even
and odd levels are
\begin{eqnarray}
\label{eq:QDDt0Even}
\lim_{T\to 0} U^{\tt QDD}_{2n}&=&1-in(\epsilon_x \sigma^x+\epsilon_y\sigma^y)\\
\label{eq:QDDt0Odd}
\lim_{T\to 0} U^{\tt QDD}_{2n-1}&=&(-1)^n \left( 1-in\epsilon_x \sigma^x \right)\;.
\end{eqnarray}
That is, for short total evolution time,
in a sequence QDD$_{2n}$, the central spin rotates
about an axis lying in the $x$-$y$ plane. Preservation
of the spin states $|X\rangle$ and $|Y\rangle$ is determined
by the relative magnitude of $\epsilon_x$, and $\epsilon_y$.
If $\epsilon_x\gg \epsilon_y$, then $|X\rangle$ is better preserved,
and vice versa.
In QDD$_{2n-1}$, the spin evolution is a rotation about the $x$-axis
by an angle $n\epsilon_x$, hence $|Y\rangle$ is better preserved.
Note that the rotation axis errors do not contribute in first order
to either even or odd level QDD for $T\to 0$.
The above features for the initial fidelity loss
are seen in Fig.~\ref{fig:QDD3QDD6}, where the simulation results
of QDD$_3$ (with 20 pulses) and QDD$_6$ (with 48 pulses) are shown for dynamical spin bath.
For QDD$_3$, at short times, $|X\rangle$ is preserved somewhat better than $|Y\rangle$,
due to the smallness of $\epsilon_x$ in our experiments.
For QDD$_6$, $|X\rangle$ and $|Y\rangle$ exhibit
the same behavior at small times, since we have $\epsilon_x=\epsilon_y$.

We did not perform analytical calculations for QDD with finite $T$
since the expansion of the evolution operator is rather cumbersome.
Compared to UDD protocol, QDD exhibits much less sensitivity
to initial states. In both QDD$_3$ and QDD$_6$, the evolutions of $S_X$
and $S_Y$ are close to those with ideal pulses.

\begin{figure}[htbp]
\includegraphics[height=4cm,angle=0]{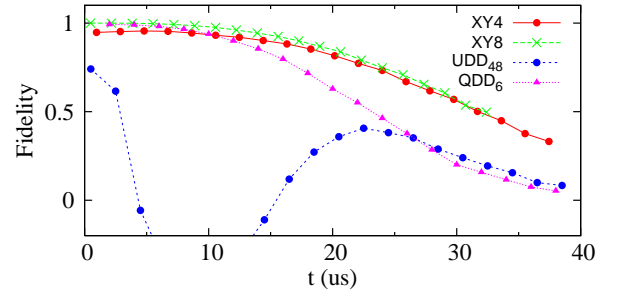}
\caption{\label{fig:QDD}
(Color online).
Fidelity $S_Y$ for the $|Y\rangle$ initial state,
for different DD sequences in the presence of pulse errors.
Results for XY4 (red line and circles), XY8 (green line and crosses), UDD$_{48}$ (blue line and dots), and QDD$_6$ (magenta line and triangles) are shown. Each sequences contains 48 pulses.
}
\end{figure}

To get a comprehensive summary of the behavior of different DD protocols in the presence of pulse errors,
Fig.~\ref{fig:QDD} shows the simulation results for 1) XY4 with 12 periods,
2) XY8 with 6 periods, 3) UDD$_{48}$, and 4) QDD$_6$.
The number of pulses is 48, the same for each sequence.
Since $S_X$ in all cases is only slightly affected by the pulse errors,
we plot only the fidelity $S_Y$ to compare different DD protocols.
The sequence XY4, which is sensitive to the in-plane rotation axis errors, $m_x$ and $n_y$, exhibits very small initial decay.
Sequence XY8, where the pulse errors are absent in the first order, exhibits no visible effect of the errors, and coincides with the curve for ideal pulses.
In the case of UDD based on $\pi_\text X$ pulses,
the state $|Y\rangle$ is rapidly destroyed by the error accumulation, as expected.
In QDD$_6$, $|Y\rangle$ is little affected by the pulse errors, see Fig.~\ref{fig:QDD3QDD6}(b),
and hence is much better preserved than in UDD$_{48}$, but its performance in the case of the O-U noise is worse in comparison with XY4 and XY8.
Moreover, the leading influencing pulse errors for different DD sequences are summarized in Table~\ref{tb:err}.

As a result, we come to the following conclusion. In the realistic case of the dynamical spin bath (O-U noise) and imperfect pulses, XY4 and XY8 sequences would be the best choices. These two have the same pulse timing, but the accumulation of pulse errors are better suppressed in the latter.

\section{Discussion of the internal structure of the spin bath composed of P1 centers}
\label{sec:bathlines}
In the above discussions, we approximated the whole spin bath as a random noise field $B(t)$. A closer examination of the spin bath
reveals that, with the static field applied along the symmetry axis of the NV center, the ESR spectrum of the P1 centers is composed of six Lorentzian lines. Thus, the NV center is decohered by six different baths, each one described via its own O-U process.
Here we demonstrate that the main conclusions drawn above remain valid.

Each P1 center is composed of an electron spin $S_k=1/2$
and a nuclear spin $I_k=1$ ($^{14}N$ nucleus) coupled via hyperfine interaction. The symmetry of this coupling, and the orientation of the nucleus' quadrupolar axis, are determined by the Jahn-Teller distortion at a given P1 location: this distortion reduces the original (local) $T_d$ symmetry of the defect to $C_{3v}$. Thus, the hyperfine and the quadrupolar Hamiltonians acquire uniaxial symmetry. The symmetry axis is oriented along one of the four $\langle 111\rangle$ directions in the diamond lattice. But the external static field applied along the $[111]$ axis separates this direction from the other three $\langle 111\rangle$ axes (which remain equivalent between themselves).
Correspondingly, all P1 centers can be categorized into two types.
Type 1 has its symmetry axis along $[111]$,
and type 2 has its symmetry axis along any of $[11\bar 1]$, $[1\bar 11]$, and $[\bar 111]$.
The Hamiltonian for a single P1 center of type 1 is\cite{Hanson08}
\begin{equation}
\label{eq:HamiP1}
{\cal H}_{\text t1} = B_0 S^z_k + A_{z} S^z_k I^z_k + A_x (S^x_kI^x_k+S^y_kI^y_k) + P (I^z_k)^2
\end{equation}
where $A_z=114~$MHz, and $A_x=81.3~$MHz are the hyperfine couplings along and perpendicular to the symmetry axis of the P1 defect, and
$P=-4~$MHz is the quadrupolar coupling constant.
The transition frequencies between different pairs of eigenstates for Hamiltonian Eq.~(\ref{eq:HamiP1})
can then be calculated. For the static field $B_0=114~$Gauss,
the flip-flops between $S_k$ and $I_k$ are greatly suppressed,
and $S^z_k$ and $I^z_k$ are good approximate quantum numbers.
The three main transitions for the electron spin with the frequencies $\nu_{1,3,6}=217,~339,~441~$MHz
correspond to $I^z_k\approx -1,~0,~1$, respectively.
Moreover, in this region of external fields, for a P1 center of type 2,
the quantization axis is also close to $z$,
and $S^z$ and $I^z$ are again good quantum numbers.
The main transitions having noticeable spectral weight, with the frequencies
$\nu_{2,4,5}=249$, 347, and 417~MHz,
also take place between the states with approximately same value of $I^z_k$.

For a single P1 center spin, all
the other surrounding P1 centers can be treated as a spin bath (made of six baths, corresponding to six spectral lines), which create
random field along the quantization axis.
Taking into account the statistical properties of
the dipolar coupling constants,\cite{KlauderAnderson,Dobrovitski08} one finds that each spectral line of the bath has Lorentian form, and is broadened due to the dipolar interaction between P1 centers by $\sim 1~$MHz.\cite{GijsBathCtrl}
%
Each spectral line can be considered as a ``spin pocket'',
and the flip-flops between the electron spins of two P1 centers can happen only if both centers belong to the same pocket, i.e.\ each spectral line creates its own noise field with rms $b_k$ and the correlation decay rate $R_k$, with $k=1,\cdots,6$.
In addition, we can include the quasi-static fluctuations in the applied magnetic field $B_0$, and describe it as another bath, with rms $b_7$ and $R_7=0$.

The contributions from these independent noise fields
to the decay factor are additive, and for the free induction decay
and the spin echo we have~\cite{KlauderAnderson}
$F(T)=\exp[-T^2\sum_{j=1}^7 {b^2_j}/2]$
and
$E(T=2\tau)=\exp[-(2\tau)^3\sum_{j=1}^6 {b^2_j R_j}/12]$, respectively.
Similarly, using the results of Sec.~\ref{sec:scaling}, for the experimentally optimal sequences XY4 and XY8 (and for the corresponding concatenated protocols) we obtain the DD fidelity
\begin{equation}
\label{eq:EtSum}
S(T)=\exp\Big[-\frac{T^3}{3N_d^2}\sum_{j=1}^7 b_j^2 R_j\Big],
\end{equation}
i.e.\ the overall effect of seven independent spin baths
is equivalent to a single noise field with the parameters $b$ and $R$ satisfying
$b^2=\sum_{j=1}^7 b^2_j$ and $b^2 R=\sum _{j=1}^7 b^2_j R_j$.
%
In a similar fashion, the main results of Sec.~\ref{sec:err} concerning the influence of the pulse errors, also remain valid for several independent baths.

\section{conclusion}
We studied dynamical decoupling (DD) protocols for the nitrogen-vacancy center in diamond, which can be used to decouple the electron spin of the NV
center from the spin bath of P1 centers.
The decoupling efficiency of various DD schemes is studied, including
periodic sequences based on single-axis and double-axis control pulses (CPMG, PDD XY, and XY4), their concatenated and symmetrized versions,
as well as aperiodical sequences UDD and QDD.

For the periodical decoupling protocol PDD XY,
the fidelity at short times decays 4 times faster than at long times. In the case of the XY4 protocol, the decay rate is uniform at short and long times, and
coincides with the slower (long-time) decay rate of PDD XY sequence.
For the spin bath under consideration,
concatenated and symmetrized versions of the periodic sequences do not improve the decoupling performance,
and the aperiodical protocols UDD and QDD exhibit inferior decoupling performance in comparison with XY4.

We also studied the effect of the accumulation of small pulse errors in the course of decoupling.
For the DD sequences based on single-axis control (CPMG and UDD),
different initial states are affected differently by the imperfect pulses, and are mainly affected by the errors
in the rotation angle and the $z$-component of the rotation axis. Thus, the single-axis protocols preserve well only one component of the central spin.
For  the sequences based on two-axis controls, XY and XY4, the errors associated with the in-plane components of the rotation axes affect the robustness of decoupling in the leading order.
In our experimental setup for the decoupling of NV center spin
from the P1 centers, such errors are small, ensuring robustness of
these protocols, and efficient preservation of all components of the central spin. Similarly, the two-axis QDD protocols are much less sensitive to the pulse errors than their single-axis counterpart UDD.

Overall, for the experimentally relevant case of the dynamical spin bath made of P1 centers, for realistically imperfect pulses, the best choices for an experimentalist are XY4 and XY8 sequences. The former is simpler, but the latter ensures better robustness with respect to the pulse errors. Concatenation does not significantly improve performance, but makes the sequences more complex.

\acknowledgements
We would like to thank M. J. Biercuk, P. Cappellaro, D. G. Cory, K. Khodjasteh, and L. Viola for useful discussions.
G.d.L., D.R., and R.H. acknowledge support from the DARPA QuEST program, the Dutch Organization for Fundamental Research on Matter (FOM), the Netherlands Organization for Scientific Research (NWO), and the EU SOLID and DIAMANT programs.
Work at Ames Laboratory was supported by the Department of Energy -- Basic Energy Sciences under Contract No DE-AC02-07CH11358.

\appendix
\section{Derivation for the decay of decoupling fidelity}
\label{app:analysis}
We consider a decoupling sequence containing $N_c$ cycles,
each of duration $T_c$ (so that $T=N_c T_c$). Each cycle
is characterized by the filter function $\xi_0(t)$
defined in Sec.~\ref{sec:analy_general}. Expressing $S(T)$ as
\begin{equation}
S(T)=\exp{[-b^2W(T)]}\;,
\end{equation}
to calculate the decay exponent $W(T)=\int_0^T ds {\rm e}^{-Rs} p(s)$, we break
the domain $[0,T]$ into $N_c$ pieces of length $T_c$:
\begin{eqnarray}
\label{eq:WT0}
W(T)&=& \sum_{m=0}^{N_c-1} \int_{m T_c}^{(m+1)T_c} {\rm e}^{-Rs} p(s) ds
\end{eqnarray}
and calculate $p(s)$ at each segment $[m T_c, (m+1) T_c]$ separately.
Expressing $s=m T_c + s'$ (with $0<s'< T_c$), and taking into
account that $\xi(t)$ and $\xi(t+s)$ overlap over $N_c-m$ full cycles,
we have
\begin{eqnarray}
p(s) &=& (N-m) [q_{11}(s')+q_{12}(s')] - q_{12}(s')
\end{eqnarray}
where
\begin{eqnarray}
q_{11}(s') &=& \int_0^{T_c-s'} \xi_0(t)\xi_0(t+s') dt \\
q_{12}(s') &=& q_{11}(T_c-s') \nonumber\;.
\end{eqnarray}
Substitute $p(s)$ into Eq.~(\ref{eq:WT0}), using ${\rm e}^{-Rs}={\rm e}^{-m R T_c} {\rm e}^{-Rs'}$,
we obtain
\begin{eqnarray}
\label{eq:WTpre}
W(T) &=& \Gamma_N \int_0^{T_c} {\rm e}^{-Rs'} [q_{11}(s')+q_{12}(s')] ds'\\\nonumber
&-&P_N \int_0^{T_c} {\rm e}^{-Rs'} q_{12}(s') ds'
\end{eqnarray}
where
\begin{eqnarray}
P_N&=&\sum_{m=0}^{N_c-1} {\rm e}^{-mRT_c}\\\nonumber
\Gamma_N&=&\sum_{m=0}^{N_c-1} (N_c-m){\rm e}^{-mRT_c}\;.
\end{eqnarray}
Here $P_N$ is a geometric progression and $\Gamma_N=N_c P_N + \frac{1}{T_c} \frac{d P_N}{dR}$.
We therefore have
\begin{eqnarray}
\label{eq:PN}
P_N&=&\frac{1-{\rm e}^{-N_cRT_c}}{1-{\rm e}^{-RT_c}}\\\nonumber
\Gamma_N &=& \frac{N_c-(N_c+1){\rm e}^{-RT_c}+{\rm e}^{-(N_c+1)RT_c}}
  {(1-{\rm e}^{-RT_c})^2}\;.
\end{eqnarray}
With the integrals in Eq.~(\ref{eq:WTpre}) expressed in terms of $Q_{11}$ and $Q_{12}$
defined in Eq.~(\ref{eq:Q11}), we obtain
\begin{equation}
\label{eq:WT}
W(T)=\big[\Gamma_N (Q_{11}+Q_{12})- P_N Q_{12}\big]\;,
\end{equation}
and the fidelity decay $S(T)$ is thus simplified to Eq.~(\ref{eq:generalST}).

The integral $Q_{12}$ can be conveniently calculated from $Q_{11}$ as
\begin{equation}
\label{eq:q12q11}
Q_{12} = {\rm e}^{-RT_c} \tilde Q_{11}
\end{equation}
where $\tilde Q_{11}$ denotes replacing $R$ by $-R$ in $Q_{11}$.
This relation will be used repeatedly
in the calculations for the fidelity decay
of specific pulse sequences.

We now calculate the decay rate $W(t)$ for different DD sequences.
focusing on the experimentally interested case  $RT_c\ll 1$,
and examine behavior at short times $RT\ll 1$, and long times $RT\gg 1$.
In theses two limits, constants $\Gamma_N$ and $P_N$ have expressions
\begin{eqnarray}
\label{eq:PGshort}
P_N &=& N_c\\\nonumber
\Gamma_N &=& N_c(N_c+1)/2\;,
\end{eqnarray}
and
\begin{eqnarray}
\label{eq:PGlong}
P_N &=& 1/(RT_c)\\\nonumber
\Gamma_N &=& N_c/(RT_c)\;,
\end{eqnarray}
respectively.

\subsection{Sequences based on PDD XY}
\label{app:XY}
In the sequence of PDD XY, the filter function has a period of $T_c=2\tau$, and:
\begin{equation}
\label{eq:xiPDD}
\xi_0(t)=\left\{
\begin{array}{cc}
+1, &\ {\rm for}\ \tau> t\ge 0\\
-1, &\ {\rm for}\ 2\tau \ge t\ge \tau \\
0, & \ {\rm otherwise}\;.
\end{array}
\right.
\end{equation}
Note that the full cycle of the PDD XY sequence, defined according to Ref.~\onlinecite{Haeberlen}, is twice longer than the period of the filter function $\xi(t)$. We can take $T_c=2\tau$, and thus significantly simplify the derivation, because we restrict ourselves to the case of pure dephasing and ideal pulses.

The convolution integral is:
\begin{equation}
q_{11}(s)=\left\{
\begin{array}{cc}
2\tau-3s, &\ {\rm for}\ \tau> s\ge 0\\
s-2\tau, &\ {\rm for}\ 2\tau > s\ge \tau \\
0, & \ {\rm otherwise}\;.
\end{array}
\right.
\end{equation}
The integrals can further be calculated as
\begin{eqnarray}
Q_{11} &=& \frac{1}{R^2}[2\delta-3+4 {\rm e}^{-\delta}-{\rm e}^{-2\delta}]\\\nonumber
Q_{12}&=& \frac{1}{R^2}[-1+4 {\rm e}^{-\delta}
-(2\delta+3) {\rm e}^{-2\delta}]\;,
\end{eqnarray}
where $\delta=R\tau$.
Focused on the experimentally interested case $\delta\ll 1$,
the decay exponent, expanded to order $\delta^4$, is
\begin{equation}
W_{\tt PDD}(T) = \frac{1}{R^2}\Big[\frac{2P_N}{3}\delta^3+\big(\frac{\Gamma_N}{3}-\frac{5P_N}{6}\big)\delta^4\Big]\;.
\end{equation}
At short times, the second term is
negligible, and the decay rate becomes
\begin{equation}
\label{eq:WPDD}
W_{\tt PDD}(T) = \frac{2N_c\delta^3}{3R^2}\;,
\end{equation}
and the fidelity decays as Eq.~(\ref{eq:SPDD}),
taking into account that $N_d=2N_c$ for PDD XY.
At long times, the decay rate is found to be
\begin{equation}
W_{\tt PDD}(T) = \frac{N_c\delta^3}{6R^2}\;,
\end{equation}
resulting in the fidelity decay as Eq.~(\ref{eq:SPDD1}).

For the concatenated sequence CDD$_{\ell}$,
consider the filter function for half a period,
we found that the integrals $Q_{11}$ and $Q_{12}$ for level $\ell$ and level $\ell-1$ are related as
\begin{equation}
\label{eq:q11cddxy}
Q^{(\ell)}_{11} = (2-{\rm e}^{-RT^{(\ell-1)}_{0c}}) Q^{(\ell-1)}_{11} - Q^{(\ell-1)}_{12}\;,
\end{equation}
where the superscripts denote the concatenation level
and $T^{(\ell-1)}_{0c}$ is the duration of the {\it full\/}
period of CDD$_{\ell-1}$.
The other integral, $Q^{(\ell)}_{12}$, can be obtained using Eq.~(\ref{eq:q12q11}).
We now derive the decay exponent for CDD$_{\ell}$  based on PDD XY.
The integrals $Q_{11}$ and $Q_{12}$ for a full period of PDD XY
can be obtained from the quantities for half a period to be
\begin{equation*}
Q^{(1)}_{11} = \frac{1}{R^2}[-7+4\delta+
12 {\rm e}^{\delta}-8 {\rm e}^{-2\delta}+4 {\rm e}^{-3\delta}-{\rm e}^{-4\delta}]
\end{equation*}
and
\begin{equation*}
Q^{(1)}_{12} = \frac{1}{R^2}[-1+4 {\rm e}^{-\delta}-8 {\rm e}^{-2\delta}+
  12 {\rm e}^{-3\delta}-(4\delta+7) {\rm e}^{-4\delta}]\;.
\end{equation*}
Substituting these result into Eqs.~(\ref{eq:q11cddxy}) and
(\ref{eq:WT}), we obtain the decay rate for CDD$_2$.
Limiting to the case $RT_c\ll 1$, we find that
\begin{eqnarray}
Q^{(2)}_{11} &=& \frac{1}{R^2}\big[\frac{8}{3} (RT_c)^3  + o(R^5\tau^5)\big] \\\nonumber
Q^{(2)}_{12} &=& \frac{1}{R^2}\big[\frac{8}{3} (RT_c)^3  (-1+RT_c) + o(R^5\tau^5)\big]\;,
\end{eqnarray}
where $T_c=8\tau$ for CDD$_2$.
Noticing Eqs.~(\ref{eq:PGshort}) and (\ref{eq:PGlong}),
we find that for both short times ($N_cRT_c\ll 1$) and for long times ($N_cRT_c\gg 1$)
the decay rate is the same, equal to the
short-time decay rate of PDD XY, Eq.~(\ref{eq:WPDD}),
taking into account that the half-period of CDD$_2$
is four times as long as the half-period of PDD XY.
This analysis can be repeated inductively, in the same manner, for any concatenation level $\ell$ (starting from CDD$_{\ell-1}$), with the same result.

Note, however, that for CDD, since $T_c$ grows exponentially with $\ell$,
the condition $RT_c\ll 1$ will be violated for higher-order CDDs,
and the corresponding analysis is not applicable anymore. However, for slow baths, the absolute value of the DD fidelity will become only smaller in this case; and for fast baths, with $RT_c\gg 1$, DD gives little improvement in any case.

\subsection{Sequences based on XY4}
\label{app:XY4}
For sequences with CPMG timing,
we take $T_c=4\tau$ and
the single-cycle filter function is
\begin{equation}
\xi_0(t)=\left\{
\begin{array}{cc}
+1, &\ {\rm for}\ \tau> t\ge 0\\
-1, &\ {\rm for}\ 3\tau > t\ge \tau\\
+1, &\ {\rm for}\ 4\tau> t\ge 3\tau\\
0, &\ {\rm others}
\end{array}
\right.
\end{equation}
The convolution integral is then:
\begin{equation}
q_{11}(s)=\left\{
\begin{array}{cc}
4\tau-5s, &\ {\rm for}\ \tau> s\ge 0\\
-s, &\ {\rm for}\ 2\tau > s\ge \tau\\
3s-8\tau, &\ {\rm for}\ 3\tau > s\ge 2\tau\\
4\tau-s, &\ {\rm for}\ 4\tau > s\ge 3\tau\\
0, &\ {\rm others}\;.
\end{array}
\right.
\end{equation}
$Q_{11}$ and $Q_{12}$ can be correspondingly calculated as
\begin{eqnarray*}
Q_{11}&=& \frac{1}{R^2}[4\delta-5+4 ({\rm e}^{-\delta}
+ {\rm e}^{-2\delta}- {\rm e}^{-3\delta})+{\rm e}^{-4\delta}]
\\
Q_{12} &=& \frac{1}{R^2}[1-4 ({\rm e}^{-\delta}
- {\rm e}^{-2\delta} - {\rm e}^{-3\delta})-(4\delta+5) {\rm e}^{-4\delta}]\;.
\end{eqnarray*}

Limiting to the case $\delta\ll 1$,
at both short and long times,
the decay rate of the fidelity is found to be the same:
\begin{equation}
\label{eq:WXY4}
W_{\tt XY4}(T) = \frac{4N_c\delta^3}{3R^2}\;.
\end{equation}
Taking into consideration that $N_d=4N_c$ for XY4,
we arrive at the scaling relation Eq.~(\ref{eq:scaling}).

For concatenated sequence CDD$^{\tt XY4}_{\ell}$,
taking $T_c$ to be half its period, we found that general
relation between integrals for levels $\ell$ and those for $\ell-1$ holds as
\begin{eqnarray}
Q^{\ell}_{11} &=&
(4- {\rm e}^{-RT^{(\ell-1)}_{0c}}-2 {\rm e}^{-2RT^{(\ell-1)}_{0c}}+ {\rm e}^{-3RT^{(\ell-1)}_{0c}})Q^{(\ell-1)}_{11} \nonumber\\
&+& (-1-2 {\rm e}^{-RT^{(\ell-1)}_{0c}}+ {\rm e}^{-2RT^{(\ell-1)}_{0c}}) Q^{(\ell-1)}_{12}\;.
\end{eqnarray}
Starting from XY4, decay rate for any higher level CDD can be derived.
Focusing on our interested case $\delta\ll 1$, the integrals for CDD$^{\tt XY4}_{\ell}$
is found to be
\begin{equation}
Q_{11}= 8 b (R\tau)^3/R^2 + o(R^5\tau^5),
\end{equation}
where the terms of order $R^4\tau^4$ are absent, and
\begin{equation}
Q_{12}= 8 b (-1+RT_c) (R\tau)^3/R^2 + o(R^5\tau^5).
\end{equation}
Substituting these integrals into Eq.~(\ref{eq:WT}),
we find that at both short and long times, the decay rates
of CDD$^{\tt XY4}_{\ell}$ and CDD$^{\tt XY4}_{\ell-1}$ are the same,
equal to the decay rate of XY4, Eq.~(\ref{eq:WXY4}),
taking into consideration the correct relation between
$T_c$ for different concatenation levels.



\begin{thebibliography}{100}
\bibitem{JelezkoGate04} F. Jelezko, T. Gaebel, I. Popa, M. Domhan, A. Gruber, J. Wrachtrup, Phys. Rev. Lett. {\bf 93}, 130501 (2004).
\bibitem{Childress06PRL} L. Childress, J. M. Taylor, A. S. Sorensen, and M. D. Lukin, Phys. Rev. Lett. {\bf 96}, 070504 (2006).
\bibitem{Dutt07}M. V. G. Dutt {\it et al.\/},
Science {\bf 316}, 1312 (2007). 
\bibitem{Cappellaro09} P. Cappellaro, L. Jiang, J. S. Hodges, and M. D. Lukin, Phys. Rev. Lett. {\bf 102}, 210502 (2009).
\bibitem{Jiang09} L. Jiang {\it et al.\/},
Science {\bf 326}, 267 (2009).
\bibitem{Neumann10} P. Neumann {\it et al.\/}, Nat. Phys. {\bf 6}, 249 (2010).
\bibitem{Fuchs11} G. D. Fuchs, G. Burkard, P. V. Klimov and D. D. Awschalom, Nat. Phys. {\bf 7}, 789 (2011). 
\bibitem{Buckley10}    B. B. Buckley,    G. D. Fuchs,    L. C. Bassett and    D. D. Awschalom, Science, {\bf 330} 1212 (2010). 
\bibitem {Togan10} E. Togan {\it et al.\/},
Nature {\bf 466}, 730 (2010). 
\bibitem{Robledo11}   L. Robledo, L. Childress, H. Bernien, B. Hensen, P. F. A. Alkemade and R. Hanson, Nature {\bf 477}, 574 (2011) 

\bibitem{Taylor08} J. M. Taylor {\it et al.\/}, Nat. Phys. {\bf 4}, 810 (2008).
\bibitem{Balasubramanian08} G. Balasubramanian {\it et al.\/}, Nature {\bf 455}, 648 (2008).
\bibitem{Maze08} J. R. Maze {\it et al.\/}, Nature {\bf 455}, 644 (2008).
\bibitem{deLangeMagnetometry10} G. de Lange, D. Rist\`e, V. V. Dobrovitski, and R. Hanson, Phys. Rev. Lett. {\bf 106}, 080802 (2011).
\bibitem{Dolde11}  F. Dolde {\it et al.\/},
Nat. Phys. {\bf 7}, 459 (2011). 

\bibitem{Meriles10} C. A. Meriles {\it et al.\/}, J. Chem. Phys. {\bf 133}, 124105 (2010),
\bibitem{Budker11} L.-S. Bouchard, V. M. Acosta, E. Bauch, and D. Budker, New J. Phys. {\bf 13}, 025017 (2011).

\bibitem{McGuinness11}L. P. McGuinness {\it et al.\/}, 
Nat. Nanotechnol. {\bf 6}, 358 (2011).
\bibitem {Cole09} J. H. Cole and L. C. L. Hollenberg, Nanotechnology {\bf 20}, 495401 (2009).
\bibitem {Hall09} L. T. Hall, J. H. Cole, C. D. Hill, and L. C. L. Hollenberg, Phys. Rev. Lett. {\bf 103}, 220802 (2009).  

\bibitem{Gruber97} A. Gruber {\it et al.\/}, Science {\bf 276}, 2012 (1997).
\bibitem{Jelezko02} F. Jelezko {\it et al.\/}, Appl. Phys. Lett. {\bf 81}, 2160 (2002).
\bibitem{Harrison04} J. Harrison, M. J. Sellars, and N. B. Manson, J. Lumin. {\bf 107}, 245 (2004).
\bibitem{Childress06} L. Childress {\it et al.\/},
Science {\bf 314}, 281 (2006).

\bibitem{Hanson08} R. Hanson, V. V. Dobrovitski, A. E. Feiguin, O. Gywat, and D. D. Awschalom, Science {\bf 320}, 352 (2008).

\bibitem{Gaebel06} T. Gaebel {\it et al.\/}, Nat. Phys. {\bf 2}, 408 (2006).
\bibitem{Fuchs09} G. D. Fuchs, V. V. Dobrovitski, D. M. Toyli, F. J. Heremans, and D. D. Awschalom, Science {\bf 326}, 1520 (2009).
\bibitem{Fuchs10} G. D. Fuchs {\it et al.\/},
Nat Phys. {\bf 6}, 668 (2010).
\bibitem{Jelezko04} F. Jelezko, T. Gaebel, I. Popa, A. Gruber, and J. Wrachtrup, Phys. Rev. Lett {\bf 92}, 076401 (2004).
\bibitem{Santori06} C. Santori {\it et al.\/},
Phys. Rev. Lett. {\bf 97}, 247401 (2006).
\bibitem{Tamarat06} P. Tamarat {\it et al.\/}, Phys. Rev. Lett. {\bf 97}, 083002 (2006).
\bibitem{Haeberlen} U. Haeberlen, {\sl High Resolution NMR in solids: Selective Averaging} (Academic, New York, 1976).
\bibitem{Viola98} L. Viola and S. Lloyd, Phys. Rev. A {\bf 58}, 2733 (1998).
\bibitem{Viola99} L. Viola, S. Lloyd, and E. Knill, Phys. Rev. Lett. {\bf 83}, 4888 (1999).
\bibitem{Khodjasteh07} K. Khodjasteh and D. A. Lidar, Phys. Rev. A {\bf 75}, 062310 (2007).
\bibitem{Biercuk09} M. J. Biercuk, H. Uys, A. P. Vandevender, N. Shiga,W. M. Itano, and J. J. Bollinger, Nature (London) {\bf 458}, 996 (2009).
\bibitem{Uys09} H. Uys, M. J. Biercuk, and J. J. Bollinger, Phys. Rev. Lett. {\bf 103}, 040501 (2009).
\bibitem{Santos05} L. F. Santos and L. Viola, Phys. Rev. A {\bf 72} 062303 (2005).

\bibitem{Uhrig07} G. S. Uhrig, Phys. Rev. Lett. {\bf 98}, 100504 (2007); New J. Phys. {\bf 10}, 083024 (2008).
\bibitem{UhrigCUDD09} G. S. Uhrig, Phys. Rev. Lett. {\bf 102}, 120502 (2009).
\bibitem{West10} J. R. West, B. H. Fong and D. A. Lidar, Phys. Rev. Lett {\bf 104}, 130501 (2010).
\bibitem{Bluhm10} H. Bluhm {\it et al.\/}, Nat. Phys. {\bf 7}, 109 (2011).
\bibitem{Morton08} J. J. L. Morton {\it et al.\/}, Nature (London) {\bf 455}, 1085 (2008).
\bibitem{Tyryshkin10} A. M. Tyryshkin {\it et al.\/},
arXiv:1011.1903v2;
\bibitem{Morton06} J. J. L. Morton {\it et al.\/}, Nat. Phys. {\bf 2}, 40 (2006);
\bibitem{Du09} J. Du, X. Rong, N. Zhao, Y. Wang, J. Yang and R. B. Liu, Nature {\bf 461}, 1265 (2009).
\bibitem{MarcusDDQD} C. Barthel, J. Medford, C. M. Marcus, M. P. Hanson, and A. C. Gossard, Phys. Rev. Lett. {\bf 105}, 266808 (2010).
\bibitem{Fraval05} E. Fraval, M. J. Sellars, and J. J. Longdell, Phys. Rev. Lett. {\bf 95}, 030506 (2005).

\bibitem{deLange10} G. de Lange, Z. H. Wang, D. Rist\`e, V. V. Dobrovitski, and R. Hanson, Science {\bf 330}, 60 (2010).
\bibitem{Ryan10} C. A. Ryan, J. S. Hodges and D.G. Cory, Phys. Rev. Lett. {\bf 105}, 200402 (2010).
\bibitem{Naydenov10} B. Naydenov {\it et al.\/},
Phys. Rev. B {\bf 83}, 081201(R) (2011).
\bibitem{Slichter} C. P. Slichter, {\sl Principles of Magnetic Resonance} (Springer, Berlin, New York, 1990).
\bibitem{Gullion90} T. Gullion, D. Baker, and M. S. Conradi, J. Magn. Reson. {\bf 89}, 479 (1990).
\bibitem{Zhang07} W. X. Zhang, V. V. Dobrovitski, L. F. Santos, L. Viola, and B. N. Harmon, Phys. Rev. B {\bf 75}, 201302(R) (2007);
    W. Zhang, N. P. Konstantinidis, V. V. Dobrovitski, B. N. Harmon, L. F. Santos, and L. Viola,
   Phys. Rev. B {\bf 77}, 125336 (2008);
\bibitem{ViolaEDD} L. Viola and E. Knill, Phys. Rev. Lett. {\bf 90}, 037901 (2003).
\bibitem{Khodjasteh05} K. Khodjasteh and D. A. Lidar, Phys. Rev. Lett. {\bf 95}, 180501 (2005).
\bibitem{Cywinski08} L. Cywinski, R. M. Lutchyn, C. P. Nave, and S. Das Sarma, Phys. Rev. B {\bf 77}, 174509 (2008).
\bibitem{Pasini10} S. Pasini and G. S. Uhrig, Phys. Rev. A {\bf 81}, 012309 (2010).
\bibitem{WYang08} W. Yang and R.-B. Liu, Phys. Rev. Lett. {\bf 101}, 180403 (2008).
\bibitem{Cywinski09} L. Cywinski, W. M. Witzel, and S. Das Sarma, Phys. Rev. B {\bf 79}, 245314 (2009).
\bibitem{Suter10} G. A. Alvarez, A. Ajoy, X. Peng, and D. Suter,  Phys. Rev. A {\bf 82}, 042306 (2010).
\bibitem{Biercuk09PRA} M. J. Biercuk, H. Uys, A. P. VanDevender, N. Shiga, W. M. Itano, and J. J. Bollinger,
Phys. Rev. A {\bf 79}, 062324 (2009).
\bibitem{Hayes11} D. Hayes, K. Khodjasteh, L. Viola, and M. J. Biercuk, arXiv:1109.6002.
\bibitem{GersteinBook} B. C. Gerstein and C. R. Dybowski, {\sl Transient Techniques in NMR of Solids} (Academic Press, Orlando, 1985).
\bibitem{MortonSPAM} J. J. L. Morton, A. M. Tyryshkin, A. Ardavan, K. Porfyrakis, S. A. Lyon, and G. A. D. Briggs, Phys. Rev. A {\bf 71}, 012332 (2005).
\bibitem{BurumEtal81} D. P. Burum, M. Linder, and R. R. Ernst, J. Mag. Res. {\bf 43}, 463 (1981).
\bibitem{ShakaEtal88} A. J. Shaka, D. N. Shykind, G. C. Chingas, and A. Pines, J. Mag. Res. {\bf 80}, 96 (1988).
\bibitem{Khaneja05} N. Khaneja, T. Reiss, C. Kehlet, T. Schulte-Herbr{\"u}ggen, and S. J. Glaser,
J. Mag. Res. {\bf 172}, 296 (2005).
\bibitem{Dobrovitski10} V. V. Dobrovitski, G. de Lange, D. Rist\`e, and R. Hanson, Phys. Rev. Lett. {\bf 105}, 077601 (2010).
\bibitem{FortunatoCoryPulses} E. M. Fortunato, M. A. Pravia, N. Boulant, G. Teklemariam, T. F. Havel, and D. G. Cory,
J. Chem. Phys. {\bf 116}, 7599 (2002).
\bibitem{CoryPulses1} N. Boulant, K. Edmonds, J. Yang, M. A. Pravia, and D. G. Cory, Phys. Rev. A {\bf 68}, 032305 (2003).
\bibitem{Sengupta05} P. Sengupta and L. Pryadko, Phys. Rev. Lett. {\bf 95}, 037202 (2005).
\bibitem{Levitt} M. H. Levitt, J. Magn. Reson, {\bf 48}, 234, (1982); Prog. NMR Spectroscopy {\bf 18}, 61 (1986).
\bibitem{Suter11} A. M. Souza, G. A. Alvarez, and D. Suter, Phys. Rev. Lett. {\bf 106}, 240501 (2011).
\bibitem{ShakaKeelerReview} A. J. Shaka and J. Keeler, Prog. NMR Spectr. {\bf 19}, 47 (1987).
\bibitem{WangJOP11} Z.-H. Wang and V. V. Dobrovitski, J. Phys. B {\bf 44}, 154004 (2011).
\bibitem{KhodjPSE} K. Khodjasteh, V. V. Dobrovitski, and L. Viola, Phys. Rev. A {\bf 84}, 022336 (2011).
\bibitem{Wang10}  Z.-H. Wang {\it et al.\/},
arXiv:1011.6417v2.
\bibitem{Wrachtrup06} J. Wrachtrup and F. Jelezko, J. Phys.: Condens. Matter {\bf 18}, S807 (2006).
\bibitem{ChildressHF} B. Smeltzer, J. McIntyre, and L. Childress, Phys. Rev. A {\bf 80}, 050302(R) (2009).
\bibitem{KlauderAnderson} J. R. Klauder and P. W. Anderson, Phys. Rev {\bf 125}, 912 (1962).
\bibitem{Kubo1} R. Kubo, M. Toda, and N. Hashitsume, {\it Statistical
 Physics II\/} (Springer, Berlin, New York, 1998).
\bibitem{HuHartmann} P. Hu and S. R. Hartmann, Phys. Rev. B {\bf 9}, 1 (1974).
\bibitem{Salikhov81} K. M. Salikhov, S. A. Dzuba, and A. M. Raitsimring, J. Mag. Res. {\bf 42}, 255 (1981).
\bibitem{Dobrovitski09} V. V. Dobrovitski, A. E. Feiguin, R. Hanson, and D. D. Awschalom, Phys. Rev. Lett. {\bf 102}, 237601 (2009).
\bibitem{deSousa03} R. de Sousa and S. Das Sarma, Phys. Rev. B {\bf 68}, 115322 (2003).
\bibitem{Witzel05} W. M. Witzel, R. de Sousa, and S. Das Sarma, Phys. Rev. B {\bf 72}, 161306(R) (2005).
\bibitem{Dobrovitski08} V. V. Dobrovitski, A. E. Feiguin, D. D. Awschalom, and R. Hanson, Phys. Rev. B {\bf 77}, 245212 (2008).
\bibitem{Saikin07} S. K. Saikin, Wang Yao, and L. J. Sham, Phys. Rev. B {\bf 75}, 125314 (2007).
\bibitem{RBLiu07} Ren-Bao Liu, Wang Yao, and L. J. Sham, New Journ. Phys {\bf 9}, 226 (2007).
\bibitem{Maze08b} J. R. Maze, J. M. Taylor, and M. D. Lukin, Phys. Rev. B {\bf 78}, 094303 (2008).
\bibitem{Cywinski10} W. M. Witzel, M. S. Carroll, A. Morello, L. Cywinski, and S. Das Sarma, Phys. Rev. Lett. {\bf 105}, 187602 (2010).
\bibitem{Cywinski11} E. Barnes, L. Cywinski, and S. Das Sarma, Phys. Rev. B {\bf 84}, 155315 (2011).
\bibitem{CoishBaugh} W. A. Coish and J. Baugh, Phys. Stat. Sol. (b) {\bf 246}, 2203 (2009).
\bibitem{CoishLoss} W. A. Coish and D. Loss, Phys. Rev. B {\bf 70}, 195340 (2004).
\bibitem{ZhangJPCM07} W. X. Zhang, N. Konstantinidis, K. A. Al-Hassanieh, and V V Dobrovitski, J. Phys.: Cond. Matter {\bf 19}, 083202 (2007).
\bibitem{vanKampen} N. G. van Kampen, {\it Stochastic Processes in Physics and Chemistry\/} (Elsevier, Amsterdam, 1981).
\bibitem{NielsenChuang} M. A. Nielsen and I. L. Chuang, {\sl Quantum Computations and Quantum Information} (Cambridge Univ. Press, Cambridge, 2002).
\bibitem{GijsBathCtrl} G. de Lange, T. van der Sar, M. S. Blok, Z.-H. Wang, V. V. Dobrovitski, and R.
Hanson, arXiv:1104.4648. (2011)






\end{thebibliography}
\end{document}